\documentclass[letterpaper,12pt]{article}
\usepackage{arxiv}
\usepackage{booktabs} % For formal tables
\usepackage[T1]{fontenc}% optional T1 font encoding
\usepackage{todonotes} %todonotes
\usepackage{url} %url handling
\usepackage{enumerate} %enumeration
\usepackage{float}
\usepackage{enumitem}
\usepackage{amsmath}
\usepackage{verbatim}
\usepackage{footnote}
\usepackage{adjustbox}
\usepackage{multirow}
\usepackage{array}
\usepackage{makecell}
\usepackage{soul}
\usepackage{xcolor}
\usepackage{lmodern}
\usepackage{graphicx}
\usepackage{comment}
\usepackage{listings}
\usepackage{lstlinebgrd}
\usepackage{expl3,xparse}
\graphicspath{{figures/}}
\usepackage{caption}
\usepackage{subcaption}

\newcolumntype{P}[1]{>{\centering\arraybackslash}p{#1}}
\newcolumntype{M}[1]{>{\centering\arraybackslash}m{#1}}

\usepackage{enumitem,amssymb}
\newlist{todolist}{itemize}{2}
\setlist[todolist]{label=$\square$,before=\sffamily}
\usepackage{pifont}

\usepackage{xcolor}

\definecolor{mygreen}{rgb}{0,0.6,0}
\definecolor{mygray}{rgb}{0.5,0.5,0.5}
\definecolor{mymauve}{rgb}{0.58,0,0.82}

\lstdefinestyle{default_listing}
{
    backgroundcolor=\color{white},           % choose the background color; you must add \usepackage{color} or \usepackage{xcolor}
    basicstyle=\scriptsize,                % the size of the fonts that are used for the code
    breakatwhitespace=false,                 % sets if automatic breaks should only happen at whitespace
    breaklines=true,                         % sets automatic line breaking
    captionpos=b,                            % sets the caption-position to bottom
    commentstyle=\color{mygreen},            % comment style
    escapeinside={(*@}{@*)},                 % if you want to add LaTeX within your code
    extendedchars=true,                      % lets you use non-ASCII characters; for 8-bits encodings only, does not work with UTF-8
    frame=tb,                                % adds a frame around the code
    float=!th,
    keepspaces=true,                         % keeps spaces in text, useful for keeping indentation of code (possibly needs columns=flexible)
    columns=flexible,
    keywordstyle=\color{blue},               % keyword style
    numbers=left,                            % where to put the line-numbers; possible values are (none, left, right)
    numbersep=5pt,                           % how far the line-numbers are from the code
    numberstyle=\scriptsize\color{mygray}, % the style that is used for the line-numbers
    rulecolor=\color{black},                 % if not set, the frame-color may be changed on line-breaks within not-black text (e.g. comments (green here))
    showspaces=false,                        % show spaces everywhere adding particular underscores; it overrides 'showstringspaces'
    showstringspaces=false,                  % underline spaces within strings only
    showtabs=false,                          % show tabs within strings adding particular underscores
    stepnumber=1,                            % the step between two line-numbers. If it's 1, each line will be numbered
    stringstyle=\color{mymauve},             % string literal style
	tabsize=2                                % sets default tabsize to 4 spaces
}

\lstdefinestyle{java}
{
    style=default_listing,
    language=Java
}

\lstdefinestyle{smali}
{
	style=default_listing,
	comment=[l]{\#},
	alsoletter={.-/},
	emph      = [1]{.method, public, constructor, .locals, .local, .param, .end},
	emphstyle = [1]{\color{blue}},
	emph      = [2]{invoke-direct, iput-boolean, return-void, iput-object, 
	move-object/16, move/16, sget-object, const/16, aput-boolean, goto/32, 
	monitor-exit, monitor-enter, goto, move-object, move-exception, throw,
	.catchall,move-result,if-nez,const/4},
    emphstyle = [2]{\color{gray}},
}

\lstnewenvironment{envcode}[1]{
    
    \lstset{
        #1
    }
}{}

\ExplSyntaxOn
\NewDocumentCommand \lstcolorlines { O{green} m }
{
 \clist_if_in:nVT { #2 } { \the\value{lstnumber} }{ \color{#1} }
}
\ExplSyntaxOff

\input{macroses.sty}

\title{Fine-grained Code Coverage Measurement in Automated Black-box Android Testing}

\author{
  Aleksandr Pilgun\thanks{Corresponding author}\\
 SnT, University of Luxembourg\\
  Luxembourg\\
  \And
  Olga Gadyatskaya\\
  SnT, University of Luxembourg\\
  Luxembourg\\
  \And
  Stanislav Dashevskyi \\
SnT, University of Luxembourg\\
Luxembourg\\
  \And
  Yury Zhauniarovich\\
  Qatar Computing Research Institute, HBKU\\
  Qatar\\
  \And
  Artsiom Kushniarou\thanks{This work has been done when Artsiom was with SnT, University of Luxembourg, Luxembourg}\\
  PandaDoc Inc.\\
  Belarus\\
}

\begin{document}

\maketitle

\begin{abstract}
    Today, there are millions of third-party Android applications. Some of these applications are buggy or even malicious. To identify such applications, novel frameworks for automated black-box testing and dynamic analysis are being developed by the Android community, including Google. Code coverage is one of the most common metrics for evaluating effectiveness of these frameworks. Furthermore, code coverage is used as a fitness function for guiding evolutionary and fuzzy testing techniques. However, there are no reliable tools for measuring fine-grained code coverage in black-box Android app testing.

    We present the Android Code coVerage Tool, \acvtool\ for short, that instruments Android apps and measures the code coverage in the black-box setting at the class, method and instruction granularities. \acvtool\ has successfully instrumented 96.9\% of apps in our experiments. It introduces a negligible instrumentation time overhead, and its runtime overhead is acceptable for automated testing tools. We show in a large-scale experiment with Sapienz, a state-of-art testing tool, that the fine-grained instruction-level code coverage provided by \acvtool\ helps to uncover a larger amount of faults than coarser-grained code coverage metrics.    
\end{abstract}

\keywords{Android \and Automated black-box testing \and Code coverage \and Third-party applications}

\section{Introduction}%
\label{sec::introduction}

Code coverage measurement is an essential element of the software development and quality assurance cycles for all programming languages and ecosystems, including Android. It is routinely applied by developers, testers, and analysts to understand the degree to which the system under test has been evaluated~\cite{ammann2016introduction}, to generate test cases~\cite{yang2009survey}, to compare test suites~\cite{gligoric2015guidelines}, and to maximize fault detection by prioritizing test cases~\cite{yoo2012regression}. In the context of Android application analysis, code coverage has become a critical metric. Fellow researchers and practitioners evaluate the effectiveness of tools for automated testing and security analysis using, among other metrics, code coverage~\cite{choudhary2015automated,mao2016sapienz,huang2015code,kong2018automated,wang2018empirical}. It is also used as a fitness function to guide application exploration in testing~\cite{mao2016sapienz,su2017guided,koroglu2018qbe}.

Unfortunately, the Android ecosystem introduces a particular challenge for security and reliability analysis: Android applications (apps for short) submitted to markets (e.g., Google Play) have been already compiled and packaged, and their source code is often unavailable for inspection, i.e., the analysis has to be performed in the \emph{black-box} setting. Measuring the code coverage achieved in this setting is not a trivial endeavor. This is why some black-box testing systems, e.g., ~\cite{sadeghi2017patdroid,choi2013guided}, use only open-source apps for experimental validation, where the source code coverage could be measured by popular tools developed for Java, such as EMMA~\cite{emma} or JaCoCo~\cite{jacoco}.

In the absence of source code, code coverage is usually measured by instrumenting the bytecode of applications~\cite{li2013bytecode}. Within the Java community, the problem of code coverage measurement at the bytecode level is well-developed and its solution is considered to be relatively straightforward~\cite{tengeri2016negative,li2013bytecode}. However, while Android applications are written in Java, they are compiled into bytecode for the register-based Dalvik Virtual Machine (DVM), which is quite different from the Java Virtual Machine (JVM). Thus, there are significant disparities in the bytecode for these two virtual machines. 

Since the arrangement of the Dalvik bytecode complicates the instrumentation process~\cite{huang2015code}, there have been so far only few attempts to track code coverage for Android applications at the bytecode level~\cite{zeng2016automated}, and they all still have limitations.  The most significant one is the \textit{coarse granularity} of the provided code coverage metric. For example, ELLA~\cite{ella}, InsDal~\cite{liu2017insdal} and CovDroid~\cite{yeh2015covdroid} measure code coverage only at at the method level. Another limitation of the existing tools is the \textit{low percentage} of successfully instrumented apps. For instance, the tools by Huang et al.~\cite{huang2015code} and Zhauniarovich et al.~\cite{zhauniarovich2015towards} support fine-grained code coverage metrics, but they could successfully instrument only 36\% and 65\% of applications from their evaluation samples, respectively. Unfortunately, such instrumentation success rates are prohibitive for these tools to be widely adopted by the Android community. Furthermore, the existing tools suffer from \textit{limited empirical evaluation}, with a typical evaluation dataset being less than 100 apps. Sometimes, research papers do not even mention the percentage of failed instrumentation attempts (e.g., \cite{liu2017insdal,cai2017droidfax,yeh2015covdroid}). %\SD{Should we put a citation here as well?}

Remarkably, in the absence of reliable fine-grained code coverage reporting tools, some frameworks integrate their own black-box code coverage measurement libraries, e.g., \cite{mao2016sapienz,song2017ehbdroid,cai2017droidfax}. However, as code coverage measurement is not the core contribution of these works, the authors do not provide detailed information about the rates of successful instrumentation, as well as other details related to the code coverage performance of these libraries.

In this paper, we present \acvtool\ -- the Android Code coVerage measurement Tool that does not suffer from the
aforementioned limitations. The paper makes the following contributions:

\begin{itemize}
	\item An approach to instrument Dalvik bytecode in its \texttt{smali} representation by inserting probes to track code coverage at the levels of classes, methods and instructions. Our approach is fully self-contained and transparent to the testing environment.
	\item An implementation of the instrumentation approach in \acvtool, which can be integrated with any testing or dynamic analysis framework. Our tool presents the coverage measurements and information about incurred crashes as handy reports that can be either visually inspected by an analyst, or processed by an automated testing environment.
	\item Extensive empirical evaluation that shows the high reliability and versatility of our approach. \begin{itemize}
	\item  While previous works~\cite{huang2015code,zhauniarovich2015towards} have only reported the number of successfully instrumented apps\footnote{For \acvtool, it is 97.8\% out of 1278 real-world Android apps.}, we also verified whether apps can be successfully executed after instrumentation. We report that \textbf{96.9\%} have been successfully executed on the Android emulator -- it is only 0.9\% less than the initial set of successfully instrumented apps. 
	\item In the context of automated and manual application testing, \acvtool\ introduces only a \textbf{negligible instrumentation time overhead}. In our experiments \acvtool\ required on average 33.3 seconds to instrument an
        app. The runtime overhead introduced by \acvtool\ is also not prohibitive. With the benchmark PassMark application~\cite{passmark}, the instrumentation code added by \acvtool\ introduces 27\% of CPU overhead, while our evaluation of executions of original and repackaged app version by Monkey~\cite{monkey} show that there is \textbf{no significant runtime overhead} for real apps (mean difference of timing 0.12 sec). 
     \item We have evaluated whether \acvtool\ reliably measures the bytecode coverage by comparing its results with those reported by JaCoCo~\cite{jacoco}, a popular code coverage tool for Java that requires the source code. Our results show that the \acvtool\ results can be \textbf{trusted}, as code coverage statistics reported by \acvtool\ and JaCoCo are highly correlated. 
    \item  By integrating \acvtool\ with Sapienz~\cite{mao2016sapienz}, an efficient automated testing framework
        for Android, we demonstrate that our tool can be \textbf{useful} as an integral part of an automated testing or security analysis
        environment. We show that fine-grained bytecode coverage metric is better in revealing crashes, while activity coverage measured by Sapienz itself shows performance comparable to not using coverage at all. Furthermore, our experiments indicate that different levels of coverage granularity can be combined to achieve better results in automated testing.   
 \end{itemize}
   \item We release \acvtool\ as an \textbf{open-source tool} to support the Android testing and analysis community. Source code and a demo video of \acvtool\ are available at \url{https://github.com/pilgun/acvtool}.
\end{itemize}

\acvtool\ can be readily used with various dynamic analysis and automated testing tools, e.g., IntelliDroid~\cite{wong2016intellidroid}, CopperDroid~\cite{tam2015copperdroid}, Sapienz~\cite{mao2016sapienz}, Stoat~\cite{su2017guided}, DynoDroid~\cite{machiry2013dynodroid}, CuriousDroid~\cite{carter2016curiousdroid} and the like, to measure code coverage. This work extends our preliminary results reported in~\cite{pilgun2018effective,dashevskyi2018influence}.  

This paper is structured as follows. We give some background information about Android applications and their code coverage measurement aspects in Section~\ref{sec:background}. The \acvtool\ design and workflow are presented in Section~\ref{sec:implementation}. Section~\ref{sec:code_instrumentation} details our bytecode instrumentation approach. In Section~\ref{sec:evaluation}, we report on the experiments we performed to evaluate the effectiveness and efficiency of \acvtool, and to assess how compliant is the coverage data reported by \acvtool\ to the data measured by the JaCoCo system on the source code. Section~\ref{sec:usefulness} presents our results on integrating \acvtool\ with the Sapienz automated testing framework, and discusses the contribution of code coverage data to bug finding in Android apps. Then we discuss the limitations of our prototype and threats to validity for our empirical findings in Section~\ref{sec:discussion}, and we overview the related work and compare \acvtool\ to the existing tools for black-box Android code coverage measurement in Section~\ref{sec:relwork}. We conclude with Section~\ref{sec:conclusions}.

\section{Background}%
\label{sec:background}

\subsection{APK Internals}%
\label{subsec:apk_internals}

Android apps are distributed as \emph{apk} packages that contain the resources files, native libraries (\texttt{*.so}), compiled code files (\texttt{*.dex}), manifest (\texttt{AndroidManifest.xml}), and developer's signature. Typical application resources are user interface layout files and multimedia content (icons, images, sounds, videos, etc.).  Native libraries are compiled C/C++ modules that are often used for speeding up computationally intensive operations.

Android apps are usually developed in Java and, more recently, in Kotlin -- a JVM-compatible language~\cite{KotlinAnnouncement}. Upon compilation, code files are first transformed into Java
bytecode files (\texttt{*.class}), and then converted into a Dalvik executable file (\texttt{classes.dex})  that can be
executed by the Dalvik/ART Android virtual machine (DVM). Usually, there is only one \code{dex} file, but Android also supports multiple \code{dex} files. Such apps are called multidex applications.

In contrast to the most JVM implementations that are stack-based, DVM is a register-based virtual machine\footnote{We
refer the interested reader to the official Android documentation about the Dalvik bytecode internals~\cite{DexFormat}
and the presentation by Bornstein~\cite{DvmInternals_Bornstein2008}.}. It assigns local variables to registers, and the DVM
instructions (opcodes) directly manipulate the values stored in the registers. Each application method  has a set of registers defined in
its beginning, and all computations inside the method can be done only through this register set. The method parameters are also a part of this set. The parameter values sent into the method are always stored in the registers at the end of method's register set.

Since raw Dalvik binaries are difficult for human understanding, several intermediate representations have been proposed that are more
analyst-friendly: \texttt{smali}~\cite{smali,google-smali} and \texttt{Jimple}~\cite{jimple}. In
this work, we work with \texttt{smali}, which is a low-level programming
language for the Android platform. \texttt{Smali} is supported by
Google~\cite{google-smali}, and it can be viewed and manipulated using, e.g., the
\textit{smalidea} plugin for the IntelliJ IDEA/Android Studio~\cite{smali}.

The Android \emph{manifest} file is used to set up various parameters of an app (e.g., whether it has been compiled with the ``debug''
flag enabled), to list its components, and to specify the set of declared and requested Android permissions. The manifest 
provides a feature that is very important for the purpose of this paper: it allows to specify the
instrumentation class that can monitor at runtime all interactions between the Android system and the app. We rely upon
this functionality to enable the code coverage measurement, and to intercept the crashes of an app and log their details.

Before an app can be installed onto a device, it must be cryptographically signed with a developer's certificate (the
signature is located under the \texttt{META-INF} folder inside an \texttt{.apk} file)~\cite{zhauniarovich2014fsquadra}. The purpose of this signature is
to establish the trust relationship between the apps of the same signature holder: for
example, it ensures that the application updates are delivered from the same developer. Still, such signatures cannot be
used to verify the authenticity of the developer of an application being installed for the first time, as other parties
can modify the contents of the original application and re-sign it with their own certificates.  Our approach relies on
this possibility of code re-signing to instrument the apps.

\subsection{Code Coverage}%
\label{subsec:code_coverage}

The notion of \emph{code coverage} refers to the metrics that help developers to estimate the portion of the source code or
the bytecode of a program executed at runtime, e.g., while running a test suite~\cite{ammann2016introduction}.
Coverage metrics are routinely used in the white-box testing setting, when the source code is available. They allow developers to estimate the relevant parts of the source code that have never been executed by a particular set of tests, thus facilitating, e.g., regression-testing and improvement of test suites.
Furthermore, code coverage metrics are regularly applied as components of fitness functions that are used for other
purposes: fault localization~\cite{tengeri2016negative}, automatic test generation~\cite{mao2016sapienz}, and test
prioritization~\cite{tengeri2016negative}.  
In particular, security testing of Android apps falls under the black-box testing category, as the source code of
third-party apps is rarely available: there is no requirement to submit the source code to Google Play. Still, Google
tests all submitted apps to ensure that they meet the security
standards\footnote{\url{https://www.android.com/security-center/}}. It is important to understand how well a third-party app has been exercised in the black-box setting, and various Android app testing tools are routinely evaluated with respect to the achieved code coverage~\cite{huang2015code,kong2018automated,choudhary2015automated,wang2018empirical}.

There exist several levels of \emph{granularity} at which the code coverage can be measured. \textit{Statement} coverage, \textit{basic block coverage}, and \textit{function (method)} coverage are very widely used. Other coverage metrics exist as well: \textit{branch}, \textit{condition}, \textit{parameter}, \textit{data-flow}, etc~\cite{ammann2016introduction}. However, these metrics are rarely used within the Android community, as they are not widely supported by the most popular coverage tools for Java and Android source code, namely JaCoCo~\cite{jacoco} and EMMA~\cite{emma}. On the other hand, the Android community often uses the \textit{activity} coverage metric, that counts the proportion of executed activities~\cite{mao2016sapienz,azim2013targeted,zeng2016automated,carter2016curiousdroid} (classes of Android apps that implement the user interface), because this metric is useful and is relatively easy to compute.

There is an important distinction in measuring the statement coverage of an app at the source code and at the
bytecode levels: the instructions and methods within the bytecode may not exactly correspond to the instructions and
methods within the original source code. For example, a single source code statement may correspond to several bytecode
instructions~\cite{DvmInternals_Bornstein2008}. Also, a compiler may optimize the bytecode so that the number of methods
is different, or the control flow structure of the app is altered~\cite{tengeri2016negative,li2013bytecode}.
It is not always possible to map the source code statements to the
corresponding bytecode instructions without having the debug information.
Therefore, it is practical to expect that the source code statement coverage cannot be reliably measured within the
black-box testing scenario, and we resort to measuring the bytecode instruction
coverage.

\section{ACVTool Design}
\label{sec:implementation}

\acvtool\ allows to \emph{measure} and \emph{analyze} the degree to which the code of a \emph{closed-source} Android app is executed during testing, and to \emph{collect crash reports} occurred during this process. We have designed the tool to be self-contained by embedding all dependencies required to collect the runtime information into the application under the test (AUT). Therefore, our tool does not require to install additional software components, allowing it to be effortlessly integrated into any existing testing or security analysis pipeline. For instance, we have tested \acvtool\ with the random UI event generator Monkey~\cite{monkey}, and we have integrated it with the Sapienz tool~\cite{mao2016sapienz} to experiment with fine-grained coverage metrics (see details in Section~\ref{sec:usefulness}). Furthermore, for instrumentation \acvtool\ uses only the instructions available on all current Android platforms. The instrumented app is thus compatible with all emulators and devices. We have tested whether the instrumented apps work using an Android emulator and a Google Nexus phone.

\begin{figure}[t!]
	\centering
	\includegraphics[width=\textwidth]{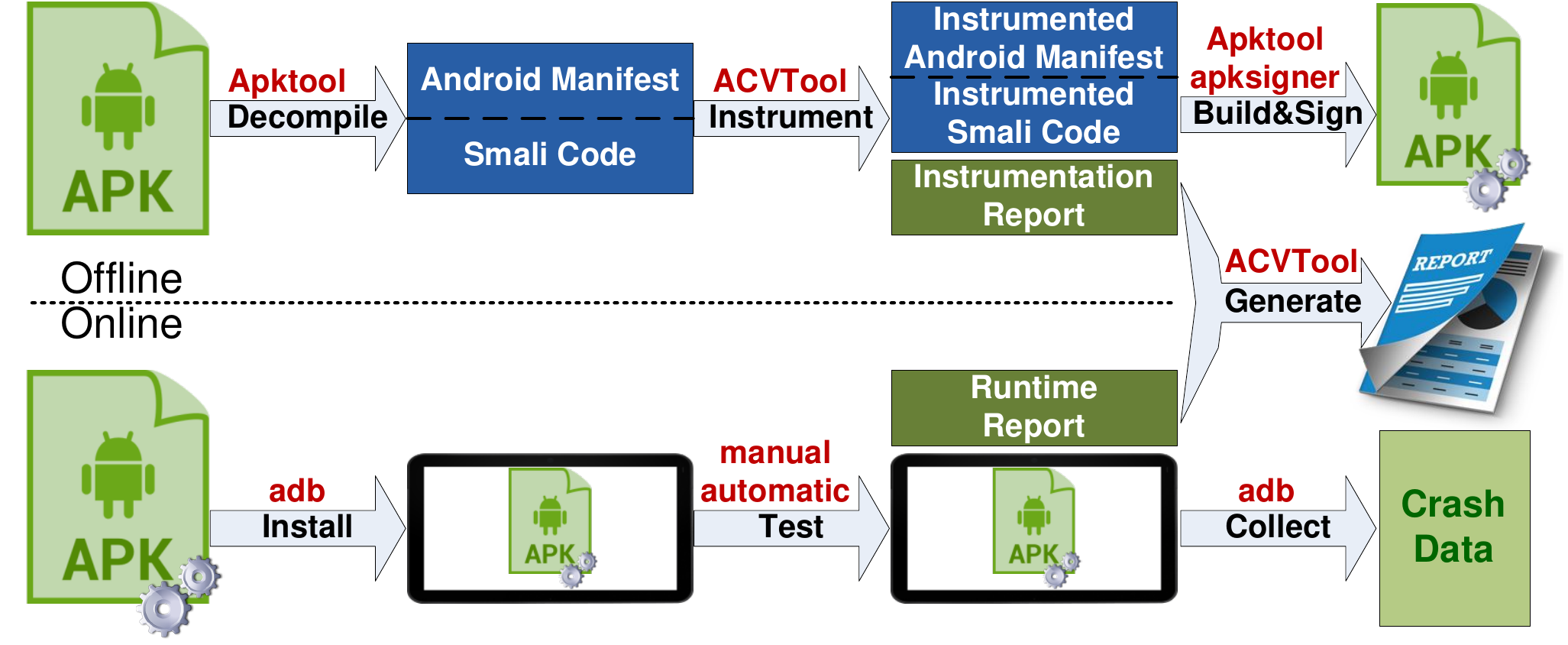}
	\caption{\acvtool\ workflow}
	\label{fig:acvtool_workflow}
\end{figure}

Figure~\ref{fig:acvtool_workflow} illustrates the workflow of \acvtool\ that consists of three phases: \textit{offline}, \textit{online} and \textit{report generation}. At the time of the offline phase, the app is instrumented and prepared for running on a device or an emulator. During the online phase, \acvtool\ installs the instrumented app, runs it and collects its runtime information (coverage measurements and crashes). At the report generation phase, the runtime information of the app is extracted from the device and used to generate a coverage report. Below we describe these phases in detail.

%%%%%%%%%%%%%%%%%%%%%%%%%%%%%%%%%%%%%%%%%%%%%%%%%%%%%%%%%%%%%%%%%%%%%%%%%%%%%%%%
\subsection{Offline Phase}
\label{subsec:offline_phase} 
The offline phase of \acvtool\ is focused on app instrumentation. In the nutshell, this process consists of several steps depicted in the upper part of Figure~\ref{fig:acvtool_workflow}. The original Android app is first decompiled using \code{apktool}~\cite{Apktool_webpage}. Under the hood, \code{apktool} uses the \code{smali/backsmali} disassembler~\cite{smali} to disassemble \texttt{.dex} files and transform them into \code{smali} representation. To track the execution of the original \code{smali} instructions, we insert special \emph{probe} instructions after each of them. These probes are invoked right after the corresponding original instructions, allowing us to precisely track their execution at runtime.  After the instrumentation, \acvtool\ compiles the instrumented version of the app using \code{apktool} and signs it with \code{apksigner}. Thus, by relying upon native Android tools and the well-supported tools provided by the community, \acvtool\ is able to instrument almost every app. We present the details of our instrumentation process in Section~\ref{sec:code_instrumentation}. 

In order to collect the runtime information, we used the approach proposed in~\cite{zhauniarovich2015towards} by developing a special \texttt{Instrumentation} class. \acvtool\ embeds this class into the app code, allowing the tool to collect the runtime information. After the app has been tested, this class serializes the runtime information (represented as a set of boolean arrays) into a binary representation, and saves it to the external storage of an Android device. The \texttt{Instrumentation} class also collects and saves the data about crashes within the AUT, and registers a broadcast receiver. The receiver waits for a special event notifying that the process collecting the runtime information should be stopped. Therefore, various testing tools can use the standard Android broadcasting mechanism to control \acvtool\ externally.

\acvtool\ makes several changes to the Android manifest file (decompiled from binary to normal xml format by \code{apktool}). First, to write the runtime information to the external storage, we additionally request the \texttt{WRITE\_EXTERNAL\_STORAGE} permission. Second, we add a special $\texttt{instrument}$ tag that registers our \texttt{Instrumentation} class as an instrumentation entry point. 

After the instrumentation is finished, \acvtool\ assembles the instrumented package with \code{apktool}, re-signs and aligns it with standard Android utilities \code{apksigner} and \code{zipalign}. Thus, the offline phase yields an instrumented app that can be installed onto a device and executed. 

It should be mentioned that we sign the application with a new signature. Therefore, if the application checks the validity of the signature at runtime, the instrumented application may fail or run with reduced functionality, e.g., it may show a message to the user that the application is repackaged and may not work properly. 

Along with the instrumented apk file, the \textit{offline} phase produces an \textit{instrumentation report}. It is a serialized code representation saved into a binary file with the \texttt{pickle} extension that is used to map probe indices in a binary array to the corresponding original bytecode instructions. This data along with the runtime report (described in Section~\ref{subsec:online_phase}) is used during the \textit{report generation} phase. Currently, \acvtool\ can instrument an application to collect instruction-, method- and class-level coverage information.

%%%%%%%%%%%%%%%%%%%%%%%%%%%%%%%%%%%%%%%

%%%%%%%%%%%%%%%%%%%%%%%%%%%%%%%%%%%%%%%%%%%%%%%%%%%%%%%%%%%%%%%%%%%%%%%%%%%%%%%%
\subsection{Online Phase}
\label{subsec:online_phase}

During the online phase, \acvtool\ installs the instrumented app onto a device or an emulator using the \texttt{adb} utility, and initiates the process of collecting the runtime information by starting the \texttt{Instrumentation} class. This class is activated through the \texttt{adb shell am instrument} command. Developers can then test the app manually, run a test suite, or interact with the app in any other way, e.g., by running tools, such as  Monkey~\cite{monkey}, IntelliDroid~\cite{wong2016intellidroid}, or Sapienz~\cite{mao2016sapienz}. \acvtool's data collection does not influence the app execution. If the \texttt{Instrumentation} class has been not activated, the app can still be run in a normal way. 

After the testing is over, \acvtool\ generates a broadcast that instructs the \texttt{Instrumentation} class to stop the coverage data collection. Upon receiving the broadcast, the class consolidates the runtime information into a \textit{runtime report} and stores it on the external storage of the testing device. Additionally, \acvtool\ keeps the information about all crashes of the AUT, including the timestamp of a crash, the name of the class that crashed, the corresponding error message and the full stack trace. By default, \acvtool\ is configured to catch all runtime exceptions in an AUT without stopping its execution -- this can be useful for collecting the code coverage information right after a crash happens, helping to pinpoint its location.

%%%%%%%%%%%%%%%%%%%%%%%%%%%%%%%%%%%%%%%%%%%%%%%%%%%%%%%%%%%%%%%%%%%%%%%%%%%%%%%%
\subsection{Report Generation Phase}
\label{subsec:report_generation_phase}
The \emph{runtime report} is a set of boolean vectors (with all elements initially set to \texttt{False}), such that each of these vectors corresponds to one class of the app. Every element of a vector maps to a probe that has been inserted into the class. Once a probe has been executed, the corresponding vector's element is set to \texttt{True}, meaning that the associated instruction has been covered. To build the \emph{coverage report}, which shows what original instructions have been executed during the testing, \acvtool\ uses data from the \emph{runtime report}, showing what probes have been invoked at runtime, and from the \emph{instrumentation report} that maps these probes to original instructions.      

\begin{figure}[t!]
	\centering
	\includegraphics[width=\columnwidth]{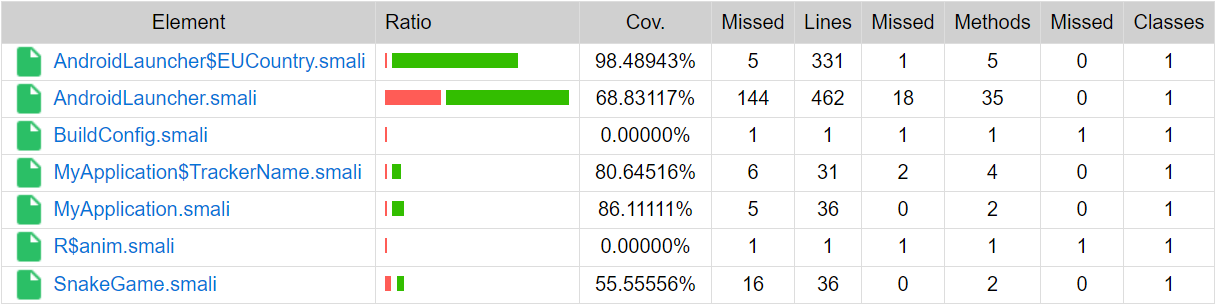}
	\caption{\acvtool\ \textit{html} report}
	\label{fig:acvtool_html_report}
\end{figure}

\begin{figure}[t!]
	\centering
	\includegraphics[width=\columnwidth]{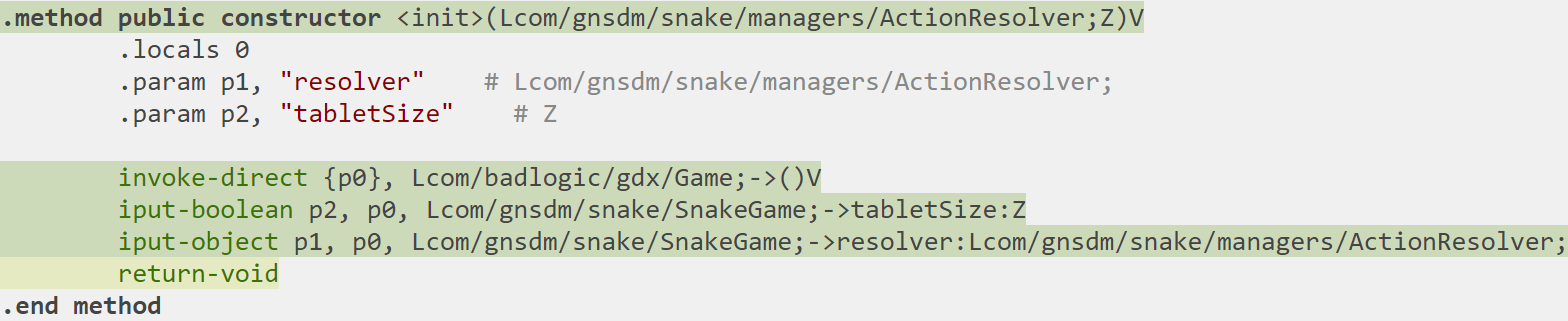}
    \caption{Covered \texttt{smali} instructions highlighted by \acvtool\ }
	\label{fig:acvtool_smali_report}
\end{figure}

Currently, \acvtool\ generates reports in the \texttt{html} and \texttt{xml} formats. These reports have a structure similar to the reports produced by the JaCoCo tool~\cite{jacoco}. While \texttt{html} reports are convenient for visual inspection, \texttt{xml} reports are more suitable for automated processing. Figure~\ref{fig:acvtool_html_report} shows an example of a \texttt{html} report. Analysts can browse this report and navigate the hyperlinks that direct to the \texttt{smali} code of individual files of the app, where the covered \code{smali} instructions are highlighted (as shown in Figure~\ref{fig:acvtool_smali_report}).

\section{Code Instrumentation} \label{sec:code_instrumentation}

Huang et al.~\cite{huang2015code} proposed two different approaches for
measuring bytecode coverage: (1) \emph{direct instrumentation} by placing
probes right after the instruction that has to be monitored for coverage (this
requires using additional registers); (2) \textit{indirect instrumentation} by
wrapping probes into separate functions. The latter instrumentation approach introduces
significant overhead in terms of added methods, that could
potentially lead to reaching the upper limit of method references per
\texttt{.dex} file (65536 methods, see~\cite{dalvik}). Thus, we built \acvtool\ upon the former approach.

\begin{envcode}{style=java, label={lst:orig-java-code}, caption={Original \textit{Java} code example.}}
private void updateElements() {
  boolean updated = false;
  while (!updated) {
    updated = updateAllElements();
  }
}
\end{envcode}

\begin{envcode}{style=smali, label={lst:orig-code}, caption={Original \textit{smali} code example.}}
.method private updateElements()V
.locals 1
  const/4 v0, 0x0
  .local v0, "updated":Z
:goto_0
  if-nez v0, :cond_0
  invoke-direct {p0}, Lcom/demo/Activity;->updateAllElements()Z (*@\label{lst:orig-code:invoke_p0}@*)
  move-result v0
  goto :goto_0
:cond_0
  return-void
.end method
\end{envcode}

\begin{envcode}{style=smali, label={lst:instr-code}, caption={Instrumented \textit{smali} code example. The yellow lines highlight the added instructions.}, linebackgroundcolor=\lstcolorlines[yellow!30]{3,4,5,6,7,9,10,12,13,15,16,19,20,23,24,26,27,28,29,30,31,32,33,34,35,36,37,38,39,40,41,42,43,44,45}}
.method private updateElements()V
.locals 4
  move-object/16 v1, p0  (*@\label{lst:instr-code:param_coping_begin}@*)
  sget-object v2, Lcom/acvtool/StorageClass;->Activity1267:[Z (*@\label{lst:instr-code:probe_args_begin}@*)
  const/16 v3, 0x1
  const/16 v4, 0x9 (*@\label{lst:instr-code:probe_args_end}@*)
  aput-boolean v3, v2, v4 (*@\label{lst:instr-code:probe_invokation}@*)
  const/4 v0, 0x0
  goto/32 :goto_hack_4
:goto_hack_back_4
:goto_0
  goto/32 :goto_hack_3
:goto_hack_back_3
  if-nez v0, :cond_0
  goto/32 :goto_hack_2
:goto_hack_back_2
  invoke-direct {v1}, Lcom/demo/Activity;->updateAllElements()Z (*@\label{lst:instr-code:call}@*)
  move-result v0   (*@\label{lst:instr-code:result}@*)
  goto/32 :goto_hack_1
:goto_hack_back_1
  goto :goto_0 (*@\label{lst:instr-code:goto}@*)
:cond_0
  goto/32 :goto_hack_0
:goto_hack_back_0
  return-void
:goto_hack_0
  const/16 v4, 0x4
  aput-boolean v3, v2, v4
  goto/32 :goto_hack_back_0
:goto_hack_1
  const/16 v4, 0x5
  aput-boolean v3, v2, v4
  goto/32 :goto_hack_back_1
:goto_hack_2
  const/16 v4, 0x6
  aput-boolean v3, v2, v4
  goto/32 :goto_hack_back_2
:goto_hack_3
  const/16 v4, 0x7
  aput-boolean v3, v2, v4
  goto/32 :goto_hack_back_3
:goto_hack_4
  const/16 v4, 0x8
  aput-boolean v3, v2, v4
  goto/32 :goto_hack_back_4
.end method
\end{envcode}

\subsection{Bytecode representation}

To instrument Android apps, \acvtool\ relies on the \code{apkil} library~\cite{Apkil_webpage} that creates a tree-based structure of \texttt{smali} code. The \code{apkil}'s tree contains classes, fields, methods, and
instructions as nodes. It also maintains relations between instructions,
labels, \texttt{try--catch} and \texttt{switch} blocks. We use this tool for two purposes: (1) \code{apkil} builds a structure representing the code that facilitates bytecode manipulations; (2) it maintains links to the inserted probes, allowing us to generate the code coverage report.

Unfortunately, \code{apkil} has not been maintained since 2013. Therefore, we
adapted it to enable support for more recent versions of Android. In
particular, we added the annotation support for classes and methods, which has
appeared in the Android API 19, and has been further extended in the API 22. We plan to support the new APIs in the future.

Tracking the bytecode coverage requires not only to insert the probes while
keeping the bytecode valid, but also to maintain the references between the
original and the instrumented bytecode. For this purpose, when we generate
the \code{apkil} representation of the original bytecode, we annotate the nodes
that represent the original bytecode instructions with additional information
about the probes we inserted to track their
execution. We then save this annotated intermediate representation of the
original bytecode into a separate serialized \texttt{.pickle} file as the instrumentation report.

\subsection{Register management}
To exemplify how our instrumentation works, Listing~\ref{lst:orig-java-code} gives an example of a Java code fragment, Listing~\ref{lst:orig-code} shows its \texttt{smali} representation, and Listing~\ref{lst:instr-code} illustrates the corresponding \texttt{smali} code instrumented by \acvtool.

The probe instructions that we insert are simple \texttt{aput-boolean} opcode instructions (e.g., Line~\ref{lst:instr-code:probe_invokation} in Listing~\ref{lst:instr-code}). These instructions put a boolean value (the first argument of the opcode instruction) into an array identified by a reference (the second argument), to a certain cell at an index (the third argument). Therefore, to store these arguments we need to allocate three additional registers per app method.

The addition of these registers is not a trivial task. We cannot simply use the first three registers in the beginning of the stack because this will require modification of the remaining method code and changing the corresponding indices of the registers. Moreover, some instructions can address only 16 registers~\cite{dalvik}, therefore the addition of new registers could make them malformed. Similarly, we cannot easily use new registers in the end of the stack because method parameters registers must always be the last ones. 

To overcome this issue, we use the following hack. We allocate three new registers, however, in the beginning of a method we copy the values of the argument registers to their corresponding places in the original method. For instance, in Listing~\ref{lst:instr-code} the instruction at Line~\ref{lst:instr-code:param_coping_begin} copies the value of the parameter \texttt{p0} into the register \texttt{v1} that has the same register position as in the original method (see Listing~\ref{lst:orig-code}). Depending on the value type, we use different \texttt{move} instructions for copying: \texttt{move-object/16} for objects, \texttt{move-wide/16} for paired registers (Android uses register pairs for \texttt{long} and \texttt{double} types), \texttt{move/16} for others. Then we update all occurrences of parameter registers through the method body from \texttt{p} names to their \texttt{v} aliases (compare the  Line~\ref{lst:orig-code:invoke_p0} in Listing~\ref{lst:orig-code} with Line~\ref{lst:instr-code:call} in Listing~\ref{lst:instr-code}). Afterwards, the last 3 registers in the stack are safe to use for the probe arguments (for instance, see Lines~\ref{lst:instr-code:probe_args_begin}-\ref{lst:instr-code:probe_args_end} in Listing~\ref{lst:instr-code}).

\subsection{Probes insertion}
Apart from moving the registers, there are other issues that must be addressed for inserting the probes correctly. First, it is impractical to insert 
probes after certain instructions that change the the execution flow of a program, namely \texttt{return}, \texttt{goto} (line~\ref{lst:instr-code:goto} in listing~\ref{lst:instr-code}), and \texttt{throw}. If a probe was placed right after these instructions, it would never be reached during the program execution. 

Second, some instructions come in pairs. For instance, the \texttt{invoke-*} opcodes, which are used to invoke a method, must be followed by the appropriate \texttt{move-result*} instruction to store the result of the method execution~\cite{dalvik} (see Lines~\ref{lst:instr-code:call}-\ref{lst:instr-code:result} in Listing~\ref{lst:instr-code}). Therefore, we cannot insert a probe between them. Similarly, in case of an exception, the result must be immediately handled. Thus, a probe cannot be inserted between the \texttt{catch} label and the \texttt{move-exception} instruction.  

These aspects of the Android bytecode mean that we insert probes after each instruction, but not after the ones modifying the execution flow, and the first command in the paired instructions. These excluded instructions are \emph{untraceable} for our approach, and we do not consider them to be a part of the resulting code coverage metric. Note that in case of a method invocation instruction, we log each invoked method, so that the computed method code coverage will not be affected by this.

The \texttt{VerifyChecker} component of the Android Runtime that checks the code validity at runtime poses additional challenges. For example, the Java \texttt{synchronized} block, which allows a particular code section to be executed by only one thread at a time, corresponds to a pair of the \texttt{monitor-enter} and \texttt{monitor-exit} instructions in the Dalvik bytecode. To ensure that the lock is eventually released, this instruction pair is wrapped with an implicit \texttt{try--catch} block, where the \texttt{catch} part contains an additional \texttt{monitor-exit} statement. Therefore, in case of an exception inside a lock, another \texttt{monitor-exit} instruction will unlock the thread. \texttt{VerifyChecker} ensures that the \texttt{monitor-exit} instruction will be executed only once, so it does not allow to add any instructions that may potentially raise an exception. To overcome this limitation, we insert the \texttt{goto/32} statement
to redirect the flow to the tracking instruction, and a label to go back after the tracking instruction was executed. Since \texttt{VerifyChecker} examines the code sequentially, and the \texttt{goto/32} statement is not considered as a statement that may throw exceptions, our approach allows the instrumented code to pass the code validity check.

\section{Evaluation}\label{sec:evaluation}

Our code coverage tracking approach modifies the app bytecode by adding probes and repackaging the original app. 
This approach could be deemed too intrusive to use with the majority of third-party applications. 
To prove the validity and the practical usefulness of our tool, we have performed an extensive empirical
evaluation of \acvtool\ with respect to the following criteria:

\textbf{Effectiveness.} We report the instrumentation success rate of \acvtool, broken down in the following numbers: 
\begin{itemize}
    \item \emph{Instrumentation success rate.} We report how many apps from our datasets have been successfully instrumented with \acvtool.
        
    \item \emph{App health after instrumentation.} We measure percentage of the instrumented apps that
        can run on an emulator. We call these apps \emph{healthy}\footnote{To the best of our knowledge, we are the first to report the percentage of
        instrumented apps that are healthy.}. To report this statistic, we installed
        the instrumented apps on the Android emulator and launched their main activity. If an app is able to run for 3 seconds without crashing, we count it as healthy.
\end{itemize}

\textbf{Efficiency.} We assess the following characteristics: 
\begin{itemize}
    \item \emph{Instrumentation-time overhead.} Traditionally, the preparation of apps for testing is considered to be
        an \emph{offline} activity that is not time-sensitive. Given that the black-box testing may be
         time-demanding (e.g., Sapienz~\cite{mao2016sapienz} tests each application for hours), our goal
        is to ensure that the instrumentation time is insignificant in comparison to the testing time. Therefore, we have measured the time \acvtool\ requires to instrument apps in our datasets.
        
    \item \emph{Runtime overhead.} Tracking instructions added into an app introduce their own runtime overhead, what may be a critical issue in testing. Therefore, we evaluate the impact of the \acvtool\ instrumentation on app performance and codebase size. We quantify runtime overhead by using the benchmark PassMark application~\cite{passmark}, by comparing executions of original and instrumented app versions, and by measuring the increase in \code{.dex} file size.
\end{itemize}

\textbf{Compliance with other tools.} We compare the coverage data reported by \acvtool\ with the coverage data measured by JaCoCo~\cite{jacoco}
which relies upon white-box approach and requires source code. This comparison allows us to draw conclusions about the reliability of the coverage information collected by \acvtool.

To the best of our knowledge, this is the largest empirical evaluation of a code coverage tool
for Android done so far. In the remainder of this section, after presenting the benchmark application sets used, we report on the results
obtained in dedicated experiments for each of the above criteria. The experiments were executed on an Ubuntu server (Xeon 4114, 2.20GHz, 128GB RAM).

\subsection{Benchmark}
We downloaded 1000 apps from the Google Play sample of the AndroZoo dataset~\cite{allix2016androzoo}. These apps
were selected randomly among apps built after Android API 22 was released, i.e., after November 2014. These are real third-party apps that may use obfuscation and anti-debugging techniques, and could be more difficult to instrument. 

Among the 1000 Google Play apps, 168 could not be launched: 12 apps were missing a launchable activity, 1 had encoding problem, and 155 that crashed upon startup. These crashes could be due to some misconfigurations in the apps, but also due to the fact that we used an emulator. Android emulators lack many features present in real devices. We have used the emulator, because we subsequently test \acvtool\ together with Sapienz~\cite{mao2016sapienz} (these experiments are reported in the next section). We excluded these unhealthy apps from the consideration. In total, our \textbf{Google Play benchmark} contains \textbf{832} healthy apps.  The apk sizes in this set range from 20KB to 51MB, with the average apk size 9.2MB.

As one of our goals is to evaluate the reliability of the coverage data collected by \acvtool\ comparing to JaCoCo as a reference, we need to have some apps
with the available source code. To collect such apps, we use the F-Droid\footnote{\url{https://f-droid.org/}} dataset of
open source Android apps (1330 application projects as of November 2017).  We managed to
\texttt{git clone} 1102 of those, and found that 868 apps used Gradle as a build system. We have successfully
compiled 627 apps using 6 Gradle versions\footnote{Gradle versions 2.3, 2.9, 2.13, 2.14.1, 3.3, 4.2.1 were used.
Note that the apps that failed to build and launch correctly are not necessarily faulty, but they can, e.g., be built
with other build systems or they may work on older Android versions. Investigating these issues is out of the scope
of our study, so we did not follow up on the failed-to-build apps.}.

To ensure that all of these 627 apps can be tested (\emph{healthy} apps), we installed them on an Android emulator and launched their main activity for 3 seconds. In total, out of these 627 apps, we obtained \textbf{446} healthy apps that constitute our
\textbf{F-Droid benchmark}. The size of the apps in this benchmark ranges from 8KB to 72.7MB, with the average size of 3.1MB.

\subsection{Effectiveness}

\subsubsection{Instrumentation success rate}
Table~\ref{tab:performance} summarizes the main statistics related to the instrumentation success rate of \acvtool.

{\footnotesize
\begin{table}[t!] \centering \caption{\acvtool\ performance evaluation}
\label{tab:performance}
\begin{tabular}{|c|c|c|c|}
\hline
\multirow{2}{*}{\textbf{Parameter}} & \textbf{Google Play}  & \textbf{F-Droid}  & \multirow{2}{*}{\textbf{Total}} \\
 & \textbf{benchmark} & \textbf{benchmark}  & \\
\hline
\hline
Total \# healthy apps & 832 & 446  & 1278 \\
\hline
Instrumented apps & 809 (97.2\%) & 442 (99.1\%)  & 1251 (97.8\%) \\
\hline
Healthy instrumented apps & 799 (96.0\%) & 440 (98.7\%)  & 1239 (96.9\%) \\
\hline
Avg. instrumentation time & 36.6 sec & 27.4 sec  & 33.3 sec \\
\hline  
\end{tabular}
\end{table}
}

Before instrumenting applications with \acvtool, we reassembled, repackaged, rebuilt (with \texttt{apktool}, \texttt{zipalign}, and \texttt{apksigner}) and installed every healthy Google Play and F-Droid app on a device. In Google Play set, one repackaged app had crashed upon startup, and \texttt{apktool} could not repackage 22 apps, raising \texttt{AndrolibException}. In the F-Droid set, \texttt{apktool} could not repackage only one app. These apps were excluded from subsequent experiments, and we consider them as failures for \acvtool\ (even though \acvtool\ instrumentation did not cause these failures).

Besides the 24 apps that could not be repackaged in both app sets, \acvtool\ has instrumented all remaining apps from the Google Play benchmark. Yet, it failed to instrument 3 apps from the F-Droid set.
The found issues were the following: in 2 cases \texttt{apktool} raised an exception \texttt{ExceptionWithContext} declaring an invalid instruction offset, in 1 case \texttt{apktool} threw \texttt{ExceptionWithContext} stating that a register is invalid and must be between \code{v0} and \code{v255}.

\subsubsection{App health after instrumentation}
From all successfully instrumented Google play apps, 10 applications crashed at launch and generated runtime exceptions, i.e., they became unhealthy after instrumentation with \acvtool\ (see the third row in Table~\ref{tab:performance}). Five cases were due absence of Retrofit annotation (four \texttt{IllegalStateException} and one \texttt{IllegalArgumentException}), 1 case -- \texttt{ExceptionInInitializerError}, 1 case -- \\ \texttt{NullPointerException}, 1 case -- \texttt{RuntimeException} in a background service. In the F-Droid dataset, 2 apps became unhealthy due to the absence of Retrofit annotation, raising \texttt{IllegalArgumentException}.

Upon investigation of the issues, we suspect that they could be due to faults in the \acvtool\ implementation. We are working to properly identify and fix the bugs, or to identify a limitation in our instrumentation approach that leads to a fault for some type of apps.  

\textbf{Conclusion:} we can conclude that \acvtool\ is able to process the vast majority of apps in our dataset, i.e., it is effective for measuring code coverage of third-party Android apps. For our total combined dataset of 1278 originally healthy apps, \acvtool\ has instrumented 1251, what constitutes 97.8\%. From the instrumented apps, 1239 are still healthy after instrumentation. This gives us the instrumentation survival rate of 99\%, and the total instrumentation success rate of 96.9\% (of the originally healthy population). The instrumentation success rate of \acvtool\ is much better than the instrumentation rates of the closest competitors BBoxTester~\cite{zhauniarovich2015towards} (65\%) and the tool by Huang et al.~\cite{huang2015code} (36\%). 

\subsection{Efficiency}
\subsubsection{Instrumentation-time overhead}
Table~\ref{tab:performance} presents the average instrumentation time required for apps from our datasets. It shows that \acvtool\ generally requires less time for instrumenting the F-Droid apps (on average, 27.4 seconds per app) than the Google Play apps (on average, 36.6 seconds). This difference is due to the smaller size of apps, and, in particular, the size of their \texttt{.dex} files.  For our total combined dataset the average instrumentation time is 33.3 seconds per app. This time is negligible comparing to the testing time usual in the black-box setting that could easily reach several hours.

\subsubsection{Runtime overhead}
\paragraph{Running two copies with Monkey} To assess the runtime overhead induced by our instrumentation in a real world setting, we ran the original and instrumented versions of 50 apps randomly chosen from our dataset with Monkey~\cite{monkey} (same seed, 50ms throttle, 250 events), and timed the executions. This experiment showed that our runtime overhead is insignificant: mean difference of timing was 0.12 sec, standard deviation 0.84 sec. While this experiment does not quantify the overhead precisely, it shows that our overhead is not prohibitive in a real-world test-case scenario. Furthermore, we have not observed any significant discrepancies in execution times, indicating that instrumented apps' behaviour was not drastically different from the original ones' behaviour, and there were no unexpected crashes. Note that in some cases two executions of the same app with the same Monkey script can still diverge due to the reactive nature of Android programs, but we have not observed such cases in our experiments.

\paragraph{PassMark overhead}
To further estimate the runtime overhead we used a benchmark application called PassMark~\cite{passmark}. Benchmark applications are designed to assess performance of mobile devices. The PassMark app is freely available on Google Play, and it contains a number of test benchmarks related to assessing CPU and memory access performance, speed of writing to and reading from internal and external drives, graphic subsystem performance, etc. These tests do not require user interaction. Research community has previously used this app to benchmark their Android related-tools (e.g.,~\cite{backes2017artist}). 

For our experiment, we used the PassMark app version 2.0 from September 2017. This version of the app is the latest that runs tests in the managed runtime (Dalvik and ART) rather than on a bare metal using native libraries. We have prepared two versions of the PassMark app instrumented with \acvtool: one version to collect full coverage information at the class, method and instruction level; and another version to log only class and method-level coverage.

\begin{table}[t!] \centering \caption{PassMark overhead evaluation}
	\label{tab:passmark}
	\begin{tabular}{|c|c|c|}
		\hline
		\multirow{2}{*}{\textbf{Granularity of instrumentation}} & \multicolumn{2}{c|}{\textbf{Overhead}} \\
		\cline{2-3}
               & \textbf{CPU} & \textbf{\texttt{.dex} size} \\
		\hline
		\hline
		Only class and method & +17\%  & +11\%\\
		\hline
		Class, method, and instruction & +27\% & +249\% \\
		\hline  
	\end{tabular}
\end{table}

Table~\ref{tab:passmark} summarizes the performance degradation of the instrumented PassMark version in comparison to the original app. When instrumented, the size of Passmark \texttt{.dex} file increased from 159KB (the original version) to 178KB (method granularity instrumentation), and to 556KB (instruction granularity instrumentation). We have run Passmark application 10 times for each level of instrumentation granularity against the original version of the app. In the CPU tests that utilize high-intensity computations, Passmark slows down, on average, by 17\% and 27\% when instrumented at the method and instruction levels, respectively. Other subsystem benchmarks did not show significant changes in numbers.

Evaluation with PassMark is artificial for a common app testing scenario, as the PassMark app stress-tests the device. However, from this evaluation we can conclude that performance degradation under the \acvtool\ instrumentation is not prohibitive, especially if it is used with modern hardware.

\paragraph{Dex size inflation}
As another metrics for overhead, we analysed how much \acvtool\ enlarges Android apps. We measured the size of \texttt{.dex} files in both instrumented and original apps for the Google Play benchmark apps. As shown in Table~\ref{tab:dex}, the \texttt{.dex} file increases on average by 157\%  when instrumented at the instruction level, and by 11\% at the method level. Among already existing tools for code coverage measurement, InsDal~\cite{liu2017insdal} has introduced \texttt{.dex} size increase of  18.2\% (on a dataset of 10 apks; average \texttt{.dex} size 3.6MB), when instrumenting apps for method-level coverage. Thus, \acvtool\ shows smaller code size inflation in comparison to the InsDal tool.

\begin{table}[t!] \centering \caption{Increase of \texttt{.dex} files for the Google Play benchmark}
	\label{tab:dex}
	\begin{tabular}{|c|c|c|c|}
		\hline
		\multirow{2}{*}{\textbf{Summary statistics}} & \multirow{2}{*}{\textbf{Original file size}} & \multicolumn{2}{c|}{\textbf{Size of instrumented file}}\\
		\cline{3-4}
   &  & \textbf{Method} & \textbf{Instruction} \\
		\hline
		\hline
		Minimum & 4.9KB & 17.6KB (+258\%) & 19.9KB (+304\%) \\
		\hline
		Median & 2.8MB & 3.1MB (+10\%)& 7.7MB (+173\%) \\
		\hline  
		Mean & 3.5MB & 3.9MB (+11\%) & 9.0MB (+157\%) \\
		\hline
		Maximum & 18.8MB & 20MB (+7\%) & 33.6MB (+78\%) \\
		\hline
	\end{tabular}
\end{table}

\textbf{Conclusion:} \acvtool\ introduces an off-line instrumentation overhead that is negligible considering the total duration of testing, which can last hours. The run-time overhead in live testing with Monkey is negligible. In the stress-testing with the benchmark PassMark app, \acvtool\ introduces 27\% overhead in CPU. The increase in code base size introduced by the instrumentation instructions, while significant, is not prohibitive. Thus, we can conclude that \acvtool\ is efficient for measuring code coverage in Android app testing pipelines.

\subsection{Compliance with JaCoCo}\label{subsec:jacoco}

\begin{figure*}[t!]
    \centering
    \begin{subfigure}[t!]{0.26\textwidth}
        \centering
        \includegraphics[width=\textwidth]{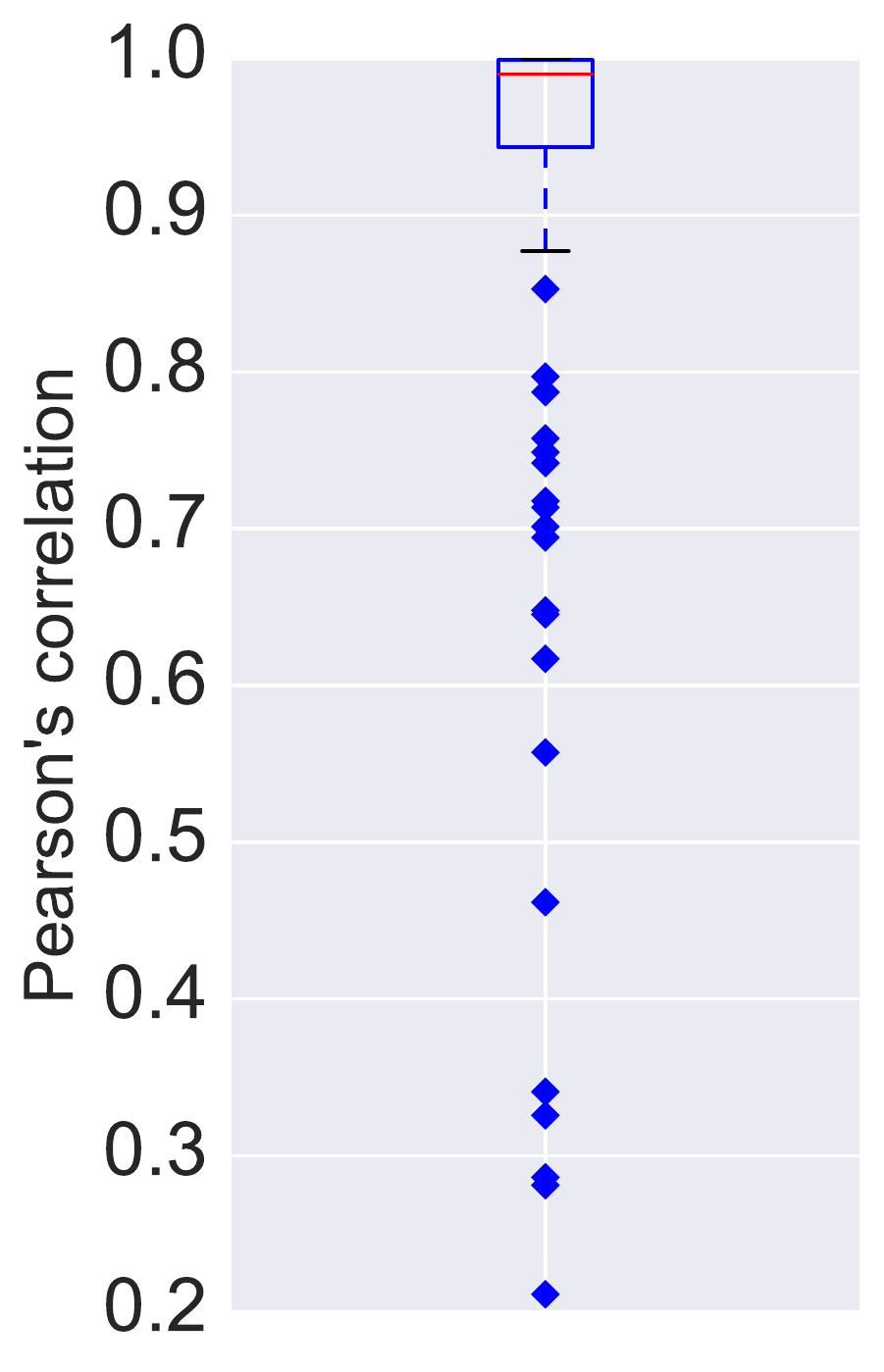}
        \caption{Boxplot of correlation for code coverage data computed by \acvtool\ and JaCoCo.}
        \label{fig-boxplot-corr}
    \end{subfigure}%
    \hspace{1cm}
    \begin{subfigure}[t!]{0.42\textwidth}
        \centering
        \includegraphics[width=\textwidth]{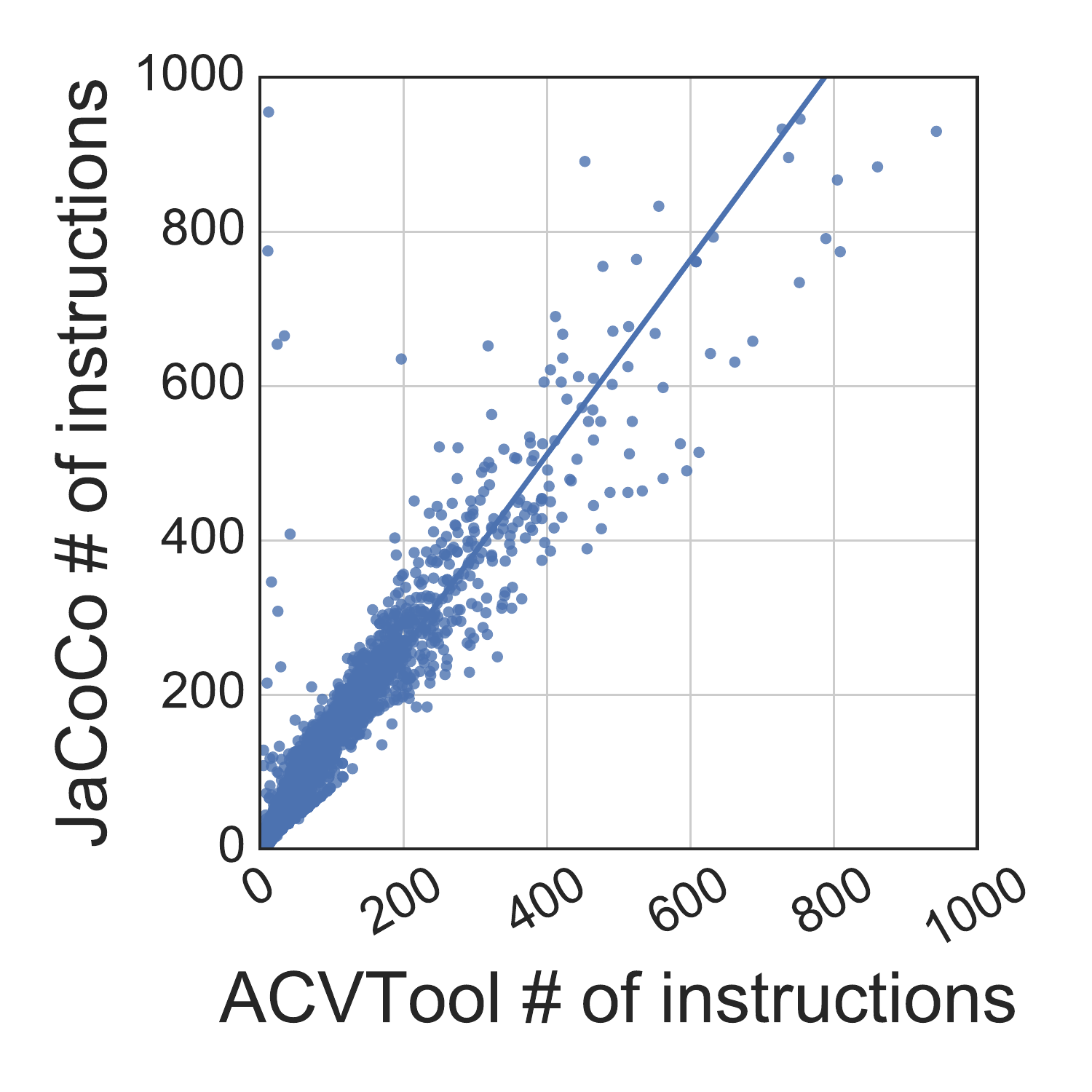}
        \caption{Scatterplot of the number of instructions in app methods, as computed by \acvtool\ and JaCoCo.}
        \label{fig-regplot}
    \end{subfigure}
\caption{Compliance of coverage data reported by \acvtool\ and JaCoCo.}
    \label{fig-compliance-data}
\end{figure*}

%[t]{0.5\textwidth}
When the source code is available, developers can log code coverage of Android apps using the JaCoCo library~\cite{jacoco} that could be integrated into the development pipeline via the Gradle plugin. We used the coverage data reported by this library to evaluate the correctness of code coverage metrics reported by \acvtool. 

For this experiment, we used only F-Droid benchmark because it contains open-source applications. We put the new \texttt{jacocoTestReport} task in the Gradle configuration file and added our \texttt{Instrumentation} class into the app source code. In this way we avoided creating app-specific tests, and could run any automatic testing tool. Due to the diversity of project structures and versioning of Gradle, there were many faulty builds. In total, we obtained 141 apks correctly instrumented with JaCoCo, i.e., we could generate JaCoCo reports for them.

We ran two copies of each app (instrumented with \acvtool\ and with JaCoCo) on the Android emulator using the same Monkey~\cite{monkey} scripts for both versions.  Figure~\ref{fig-boxplot-corr} shows the boxplot of correlation of code coverage measured by \acvtool\ and JaCoCo. Each data point corresponds to one application, and its value is the Pearson correlation coefficient between percentage of executed code, for all methods included in the app. The minimal correlation is 0.21, the first quartile is 0.94, median is 0.99, and maximal is 1.00. This means that for more than 75\% of apps in the tested applications, their code coverage measurements have correlation equal to 0.94 or higher, i.e., they are strongly correlated. Overall, the boxplot demonstrates that code coverage logged by \acvtool\ is strongly correlated with code coverage logged by JaCoCo.

The boxplot in Figure~\ref{fig-boxplot-corr} contains a number of outliers. Discrepancies in code coverage measurement have appeared due to several reasons. First of all, as mentioned in Section~\ref{sec:code_instrumentation}, \acvtool\ does not track some instructions. It is our choice to not count those instructions towards \emph{covered}. In our F-Droid dataset, about half of app methods consist of 7 \textit{smali} instructions or less. Evidently, the correlation of covered instructions for such small methods can be perturbed by these untraceable instructions.

The second reason for the slightly different code coverage reported is the differences in the \texttt{smali} code and Java bytecode. Figure~\ref{fig-regplot} shows a scatterplot of method instruction numbers in \texttt{smali} code (measured by \acvtool, including the ``untrackable'' instructions) and in Java code (measured by JaCoCo). Each point in this Figure corresponds to an individual method of one of the apks. The line in the Figure is the linear regression line. The data shape demonstrates that the number of instructions in \texttt{smali} code is usually slightly smaller than the number of instructions in Java bytecode. 

Figure~\ref{fig-regplot} also shows that there are some outliers, i.e., methods that have low instruction numbers in \texttt{smali}, but many instructions in Java bytecode. We have manually inspected all these methods and found that outliers are constructor methods that contain declarations of arrays. \texttt{Smali} (and Dalvik VM) allocates such arrays with only one pseudo-instruction (\texttt{.array-data}), while Java bytecode is much longer~\cite{DvmInternals_Bornstein2008}. 

\textbf{Conclusion:} overall, we can summarize that code coverage data reported by \acvtool\ generally agree with data computed by JaCoCo. The discrepancies in code coverage appear due to the different approaches that the tools use, and the inherent differences in the Dalvik and Java bytecodes.

\section{Contribution of Code Coverage Data to Bug Finding}
\label{sec:usefulness}

To assess the usefulness of \acvtool\ in practical black-box testing and analysis scenarios, we integrated \acvtool\ with Sapienz~\cite{mao2016sapienz} -- a state-of-art automated Android search-based testing tool. Its fitness function looks for Pareto-optimal solutions using three criteria: code coverage, number of found crashes and the length of a test suite. This experiment had two main goals: (1) ensure that \acvtool\ fits into a real automated testing/analysis pipeline; (2) evaluate whether fine-grained code coverage measure provided by \acvtool\ can be useful to automatically uncover diverse types of crashes with black-box testing strategy. 

Sapienz integrates three approaches to measure code coverage achieved by a test suite: EMMA~\cite{emma} (reports source code statement coverage); ELLA~\cite{ella} (reports method coverage); and its own plugin to measure coverage in terms of launched Android activities. EMMA does not work without the source code of apps, and thus in the black-box setting only ELLA and own Sapienz plugin could be used. The original Sapienz paper~\cite{mao2016sapienz} has not evaluated the impact of code coverage metric used on the discovered crashes population.

Our previously reported experiment with JaCoCo suggests that \acvtool\  can be used to replace EMMA, as the coverage data reported for Java instructions and \texttt{smali} instructions are highly correlated and comparable. Furthermore, \acvtool\ integrates capability to measure coverage in terms of classes and methods, and thus it can also replace ELLA within the Sapienz framework. Note that the code coverage measurement itself does not interfere with the search algorithms used by Sapienz.

As our dataset, we use the healthy instrumented apks from the Google Play dataset described in the previous section. We have run Sapienz against each of these 799 apps, using its default parameters. Each app has been tested using the activity coverage provided by Sapienz, and the method and instruction coverage supplied by \acvtool. Furthermore, we also ran Sapienz without coverage data, i.e., substituting coverage for each test suite as $0$. 

On average, each app has been tested by Sapienz for 3 hours for each coverage metric. After each run, we collected the crash information (if any), which included the components of apps that crashed and Java exception stack traces.

In this section we report on the results of crash detection with different coverage metrics and draw conclusions about how different coverage metrics contribute to bug detection. 

\subsection{Descriptive statistics of crashes}

\begin{table}[t!] 
\centering \caption{Crashes found by Sapienz in 799 apps}
\label{tab:sapienz:crashes}
\begin{tabular}{|c|c|c|c|}
\hline
    \textbf{Coverage metrics} & 
    \begin{tabular}[x]{@{}c@{}}\textbf{\# unique}\\\textbf{crashes}\end{tabular} & 
    \begin{tabular}[x]{@{}c@{}}\textbf{\# faulty}\\\textbf{apps}\end{tabular} & 
    \begin{tabular}[x]{@{}c@{}}\textbf{\# crash}\\\textbf{types}\end{tabular} \\
\hline
\hline
    Activity coverage & 547 (47\%) & 381 & 30 \\
\hline
    Method coverage & 578 (50\%) & 417 & 31 \\
\hline
    Instruction coverage & 612 (53\%) & 429 & 30 \\
 \hline
Without coverage & 559 (48\%) & 396 & 32 \\
\hline
\hline
    Total & 1151 & 574 & 37 \\
\hline
\end{tabular}
\end{table}

\begin{figure*}[t!]
    \centering
    \begin{subfigure}[t!]{0.54\textwidth}
        \centering
        \includegraphics[width=\textwidth]{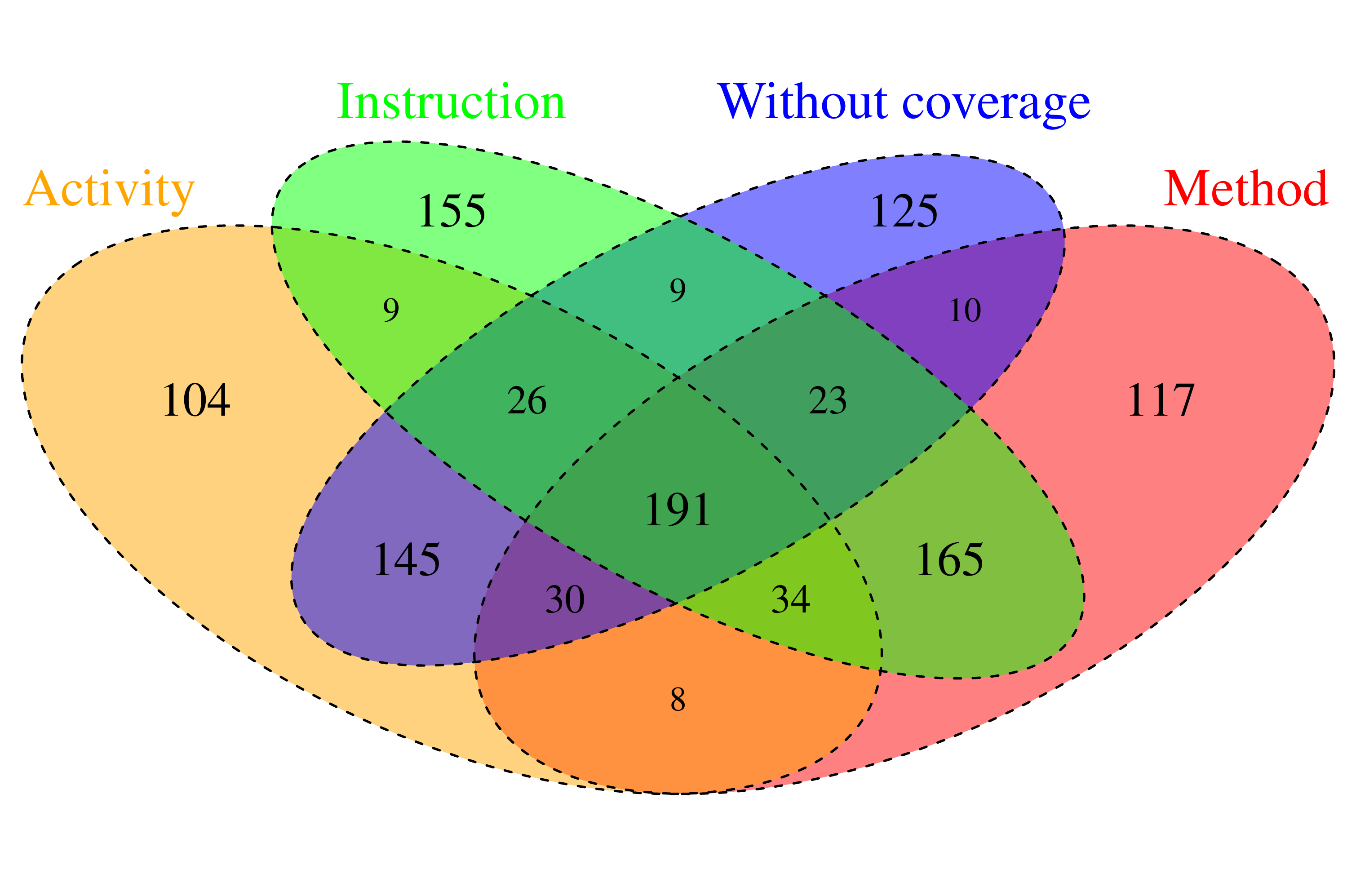}
        \caption{Crashes found with Sapienz using different coverage metrics in 799 apps.}
        \label{fig:venn}
    \end{subfigure}%
    \hspace{0.4cm}
    \begin{subfigure}[t!]{0.38\textwidth}
        \centering
        \includegraphics[width=\textwidth]{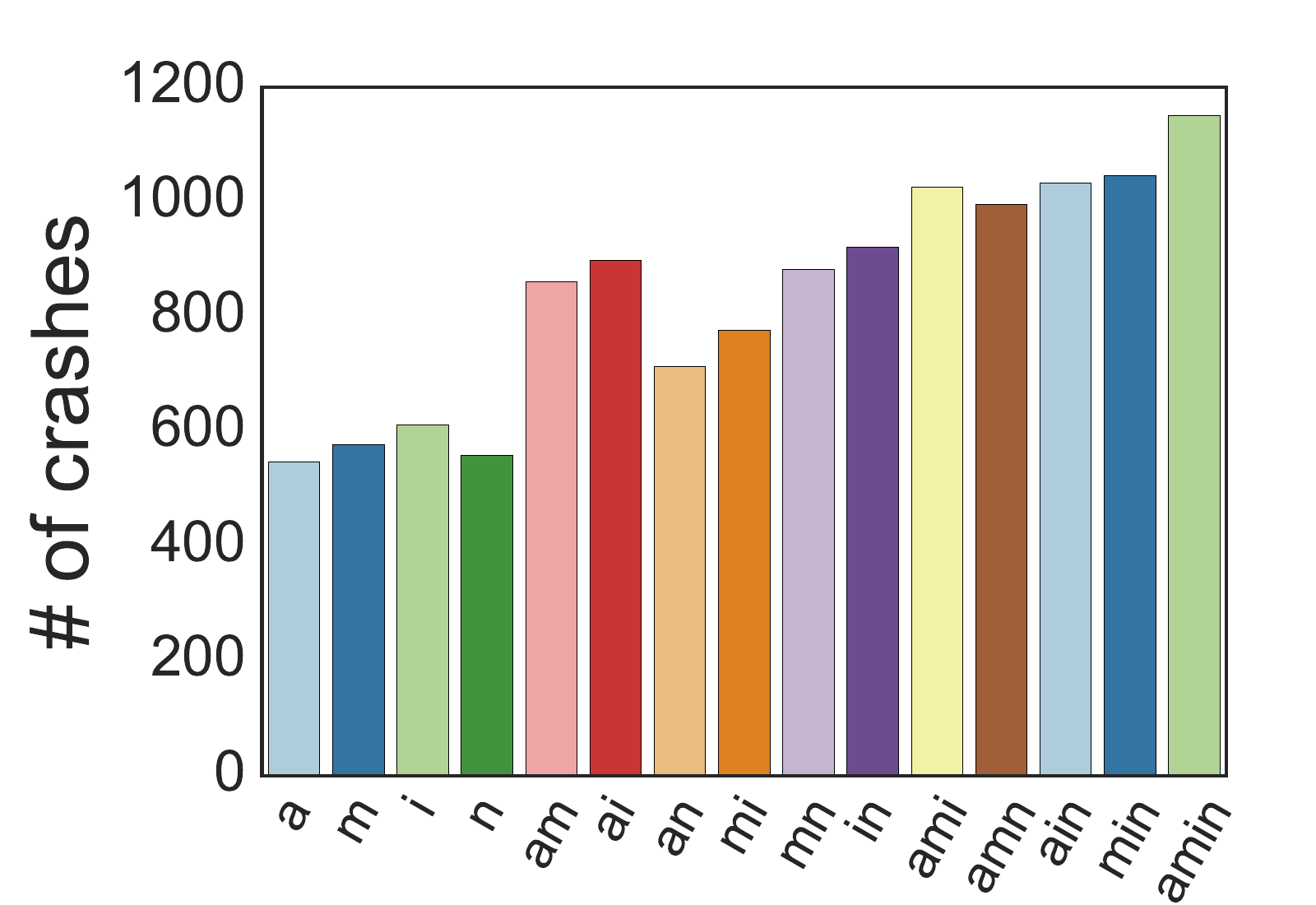}
        \caption{Barplot of crashes found by coverage metrics individually and jointly ($a$ stands for activity, $m$ for method, $i$ for instruction coverage, and $n$ for no coverage).}
        \label{fig-barplot}
    \end{subfigure}
\caption{Crashes found by Sapienz.}
    \label{fig-crashes-sapienz}
\end{figure*}

Table~\ref{tab:sapienz:crashes} shows the numbers of crashes grouped by coverage metrics that Sapienz has found in
the 799 apps. We consider a \emph{unique crash} as a unique combination of an application, its component where a crash occurred, the
line of code that triggered an exception, and a specific Java exception type.

In total, Sapienz has found 574 apps out of 799 to be faulty (at least one crash detected), and it has logged 1151 unique crashes with the four coverage conditions. Figure~\ref{fig:venn} summarizes the crash distribution for the coverage metrics. The intersection of all code coverage conditions' results contains 191 unique crashes (18\% of total crash population). Individual coverage metrics have found 53\% (instruction coverage), 50\% (method coverage), 48\% (without coverage), and 47\% (activity coverage) of the total found crashes. 

Our empirical results suggest that coverage metrics at different granularities can find distinct crashes. Particularly, we note the tendencies for instruction and method coverage, and for activity coverage and Sapienz without coverage data to find similar crashes. This result indicates that activity coverage could be too coarse-grained and comparable in effect to ignoring coverage data at all. Instruction and method-level coverage metrics, on the contrary, are relatively fine-grained and are able to drive the genetic algorithms in Sapienz towards different crash populations. Therefore, it is possible that a combination of a coarse-grained metric (or some executions without coverage data) and a fine-grained metric measured by \acvtool\ could provide better results in testing with Sapienz. 

We now set out to investigate how multiple runs affect detected crashes, and whether a combination of coverage metrics could detect more crashes than a single metric. 

\subsection{Evaluating behavior on multiple runs}
Before assessing whether a combination of metrics could be beneficial for evolutionary testing, we look at assessing the impact of randomness on Sapienz' results. Like many other automated testing tools for Android, Sapienz is non-deterministic. Our findings may be affected by this. To determine the impact of coverage metrics in finding crashes \emph{on average}, we need to investigate how crash detection behaves in multiple runs. Thus, we have performed the following two experiments on a set of 150 apks randomly selected from the 799 healthy instrumented apks. 

\subsubsection{Performance in 5 runs}
We have run Sapienz for 5 times with each coverage metric and without coverage data, for each of 150 apps. This gives us two crash populations: $\crashtotal$ that contains unique crashes detected in the 150 apps during the first experiment, and $\crashrep$ that contains unique crashes detected in the same apps running Sapienz 5 times. Table~\ref{tab:crash-pop} summarizes the populations of crashes found by Sapienz with each of the coverage metrics and without coverage.

As expected, running Sapienz multiple times increases the amount of found crashes. In this experiment, we are interested in the proportion of crashes contributed by coverage metrics individually. If coverage metrics are interchangeable, i.e., they do not differ in capabilities of finding crashes, and they will, eventually, find the same crashes, the proportion of crashes found by individual metrics to the total crashes population can be expected to significantly increase: each metric, given more attempts, will find a larger proportion of the total crash population. 

As shown in Table~\ref{tab:crash-pop}, the activity coverage has found a significantly larger proportion of total crash population (52\% from 42\%). Sapienz without coverage data also shows better performance over multiple runs (51\% from 43\%). Yet, the instruction coverage has only slightly increased performance (54\% from 50\%), while the method coverage has fared worse (49\% from 51\%). These findings suggest that the coverage metrics are not interchangeable, and even with 5 repetitions they are not able to find all crashes that were detected by other metrics. Our findings in this experiment are consistent with the previously reported smaller experiment that involved only 100 apps (see \cite{dashevskyi2018influence} for more details).

\begin{table}[t!] 
    \centering \caption{Crashes found in 150 apps with 1 and 5 runs}
    \label{tab:crash-pop}
    \begin{tabular}{|c|c|c|}
    \hline
       \multirow{2}{*}{\textbf{Coverage metrics}} & \multicolumn{2}{c|}{\textbf{Crashes}}\\
       \cline{2-3}
       & $\crashtotal$: 1 run & $\crashrep$: 5 runs \\
       \hline
       \hline
       Activity coverage & 86 (42\%) & 184 (52\%)  \\
       \hline
       Method coverage & 104 (51\%) & 174 (49\%) \\
       \hline
       Instruction coverage & 103 (50\%) & 190 (54\%) \\
       \hline
       No coverage & 89 (43\%) & 180 (51\%) \\
       \hline
       \hline
       Total & 203 & 351 \\
\hline
\end{tabular}
\end{table}

\subsubsection{The Wilcoxon signed-rank test}

The previous experiment indicates that even repeating the runs multiple times does not allow any of the code coverage metrics to find the same amount of bugs as all metrics together. The instruction coverage seems to perform slightly better than the rest in the repeating runs, but not a lot. We now fix the time that Sapienz spends on each apk\footnote{In these testing scenarios, Sapienz spends the same amount of time per app (3 runs), but the coverage conditions are different.}, and we want to establish whether the amount of crashes that Sapienz can find in an apk with 3 metrics is greater than the amount of crashes found with just one metric but with 3 attempts. This will suggest that the combination of 3 metrics is more effective in finding crashes than each individual metric. For each apk from the chosen 150 apps, we compute the number of crashes detected by Sapienz with each of the three coverage metrics executed once. We then have executed Sapienz 3 times against each apk with each coverage metric individually.

\begin{figure}[t!]
	\centering
	\includegraphics[width=0.4\columnwidth]{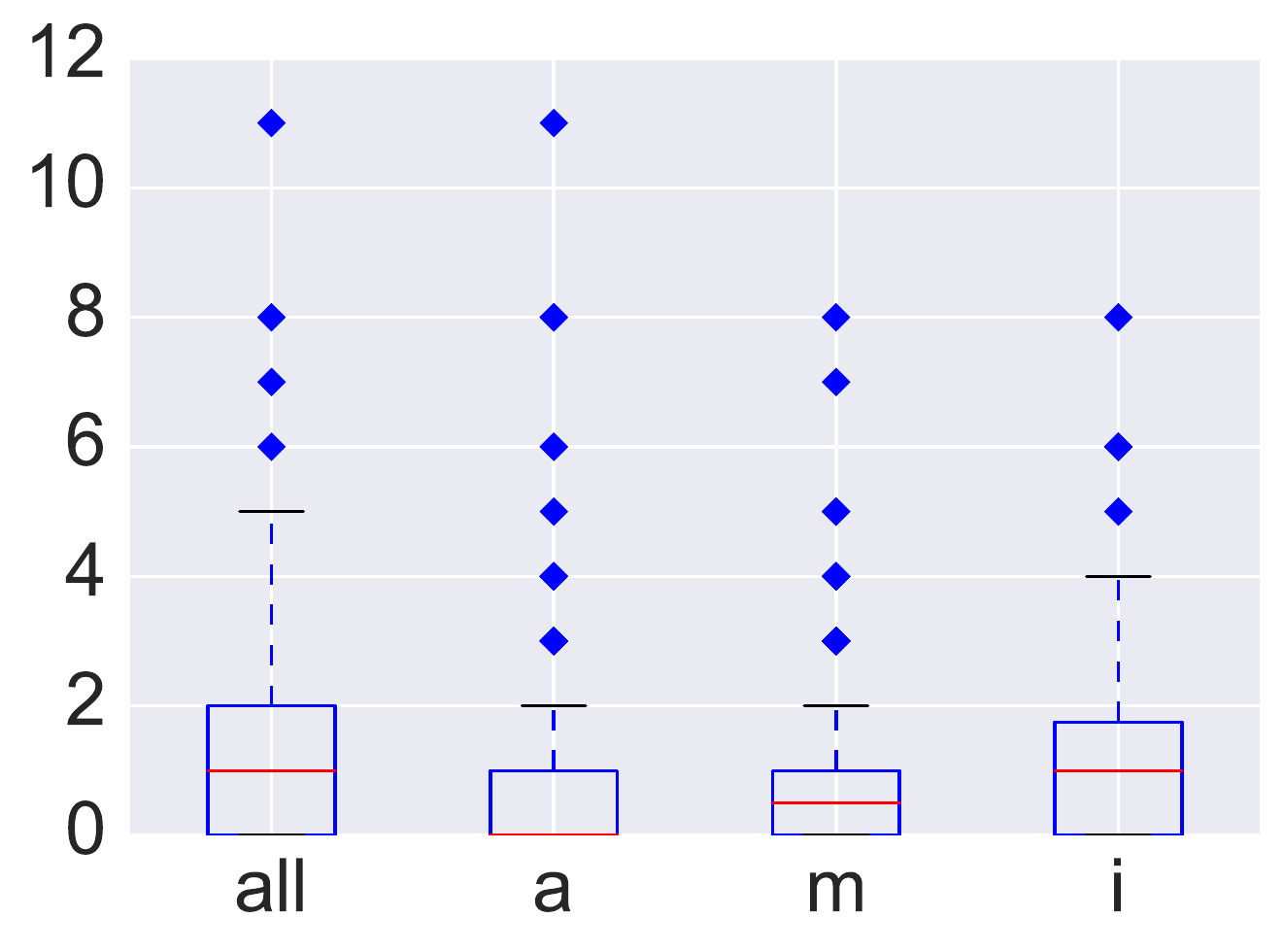}
	\caption{Boxplots of crashes detected per app ($a$ stands for activity, $m$ for method, and $i$ for instruction, respectively).}
	\label{fig:boxplot}
\end{figure}

\begin{table}[t!] 
\centering \caption{Summary statistics for crashes found per apk, in 150 apk}
\label{tab:crash-summary}
\begin{tabular}{|c|c|c|c|c|}
\hline
   \multirow{2}{*}{\textbf{Statistics}} & \multirow{2}{*}{\textbf{1 run $\times$ 3 metrics}} & \multicolumn{3}{c|}{\textbf{3 runs $\times$ 1 metric}}\\
   \cline{3-5}
   &  & activity & method & instruction \\
   \hline
   \hline
Min & 0 & 0 & 0 & 0 \\
\hline
1st. Quartile & 0 & 0 & 0 & 0 \\
\hline
Mean & 1.20 & 1.02 & 0.97 & 1.08 \\
%\hline
%St. deviation & 1.62 & 1.69 & 1.39 & 1.51 \\
\hline 
Median & 1 & 0 & 0.5 & 1\\
\hline
3rd. Quartile & 2 & 1 & 1 & 1.75 \\
\hline
Max &  11 & 11 & 8 & 8 \\
\hline  
\end{tabular}
\vspace{-8pt}
\end{table}

Table~\ref{tab:crash-summary} summarizes the basic statistics for the apk crash numbers data, and the data shapes are shown as boxplots in Figure~\ref{fig:boxplot}. The summary statistics show that Sapienz equipped with 3 coverage metrics has found, on average, more crashes per apk than Sapienz using only one metric but executed 3 times. To verify this, we apply the Wilcoxon signed-rank test~\cite{wohlin2012experimentation}. This statistical test is appropriate, as our data is paired but not necessarily normally distributed. 

The \emph{null-hypothesis} for the Wilcoxon test is that there is no difference which metric to use in Sapienz. Then, on average, Sapienz with 3 metrics will find the same amount of crashes in an app as Sapienz with 1 metric but run 3 times.  \emph{Alternative hypothesis} is that Sapienz with several coverage metrics will consistently find more crashes. Considering the standard significance level of 5\%, on our data, the results of the Wilcoxon test rejected the null-hypothesis for the activity and method coverage metrics ($p$-values 0.002 and 0.004, respectively), but not for the instruction coverage ($p$-value 0.09, which is more than the threshold 0.05). Cohen's $d$ for effect size are equal, respectively, 0.297, 0.282 and 0.168 for each of the metrics.

These results show that it is likely that a combination of coverage metrics achieves better results than only activity or method coverage. The outcome for the instruction coverage is inconclusive. We posit that a larger-size experiment could provide a more clear picture.

Indeed, our findings from this last experiment are also not fully consistent with the previously reported experiment on a smaller set of 100 apps \cite{dashevskyi2018influence}. The difference could be explained by two factors. First, we have used only healthy instrumented apps in this experiment (the ones that did not crash upon installation). The experiment reported in \cite{dashevskyi2018influence} did not involved the check for healthiness, and the crashing apps could have affected the picture. In the unhealthy app case, Sapienz always reports one single crash for it, irrespectively of which coverage metrics is used. Note that in our Google Play sample approximately 17\% are unhealthy, i.e., they cannot be executed on an emulator, as required by Sapienz. Second, the new apps tested in this experiment could have behaved slightly differently than the previously tested cohort, and the instruction coverage was able to find more bugs in them.

\subsection{Analysis of results}
Our experiments show that \acvtool\ can be integrated into an automated testing pipeline and it can be used in conjunction with available testing tools such as Sapienz. Furthermore, our results establish that a fine-grained code coverage measurement tool, such as \acvtool, can be helpful in improving automated testing tools that rely on code coverage. Indeed, in our experiment with 799 Google Play apps, the instruction-level coverage has been able to identify more faults in a single Sapienz run than other considered coverage metrics. Moreover, we compared three coverage metrics executed once with individual metrics executed 3 times, and the method and activity coverage metrics were found less effective by the Wilcoxon test of means, while the instruction coverage alone could be reasonably effective in finding bugs. 

 We can also conclude that better investigation and integration of different coverage granularities is warranted in the automated Android testing domain, just like in software testing in general~\cite{chekam2017empirical}. In our experiment with 799 apps, Sapienz without coverage data has shown results most comparable to Sapienz equipped with activity coverage. This finding could indicate that activity coverage is too coarse-grained for being a useful fitness-function in automated Android testing. On the other hand, our experiment with repeating executions 5 times shows that no coverage metrics is able to find the vast majority of the total found crash population. This result indicates that different granularities of coverage are not directly interchangeable. Further investigation of these aspects could be a promising line of research.

\section{Discussion}%
\label{sec:discussion}

\acvtool\ addresses the important problem of measuring code coverage of closed-source Android apps. Our experiments show that the proposed instrumentation approach works for the majority of Android apps, the measured code coverage is reliable, and the tool can be integrated with security analysis and testing tools. We have already shown that integration of the coverage feed produced by our tool into an automated testing framework can help to uncover more application faults. Our tool can further be used, for example, to compare code coverage achieved by dynamic analysis tools and to find suspicious code regions. 

In this section, we discuss limitations of the tool design and current implementation, and summarize the directions in which the tool can be further enhanced. We also review threats to validity regarding the conclusions we make from the Sapienz experiments.

\subsection{Limitations of \acvtool}
\acvtool\ design and implementation have several limitations. An inherent limitation of our approach is that the apps must be first instrumented before their code coverage can be measured. Indeed, in our experiments, there was a fraction of apps that could not be instrumented. Furthermore, apps can employ various means to prevent repackaging, e.g., they can check signature at the start, and stop executing in case of a failed signature check. This limitation is common to all tools that instrument applications (e.g.,~\cite{zhauniarovich2015towards,ella,huang2015code,yeh2015covdroid,liu2017insdal}). Considering this, \acvtool\ has successfully instrumented 96.9\% of our total original dataset selected randomly from F-Droid and Google Play. Our instrumentation success rates are significantly higher than any of the related work, where this aspect has been reported (e.g.,~\cite{huang2015code,zhauniarovich2015towards}). Therefore, \acvtool\ is highly practical and reliable. We examine the related work and compare \acvtool\ to the available tools in the subsequent Section~\ref{sec:relwork}.

We investigated the runtime overhead introduced due to our instrumentation, which could be another
potential limitation. Our results show that \acvtool\ does not introduce a prohibitive runtime overhead. For example, the very resource-intensive computations performed by the PassMark benchmark app degrade by 27\% in the instruction-level instrumented version. This is a critical scenario, and the overhead for an average app will be much smaller, what is confirmed by our experiments with Monkey.

We assessed that the code coverage data from \acvtool\ is compliant to the measurements from the well-known JaCoCo~\cite{jacoco} tool. We have found that, even though there could be slight discrepancies in the number of instructions measured by JaCoCo and \acvtool, the coverage data obtained by both tools is highly correlated and commensurable. Therefore, the fact that \acvtool\ does not require the source code makes it, in contrast to JaCoCo, a very promising tool for simplifying the work of Android developers, testers, and security specialists.

One of the reasons for the slight difference in the JaCoCo and \acvtool\ measurements of the number of instructions is the fact that we do not track several instructions, as specified in Section~\ref{sec:code_instrumentation}. Technically, nothing precludes us from adding probes right before the ``untraceable'' instructions. However, we consider this solution to be inconsistent from the methodological perspective, because we deem the right place for a probe to be right after the executed instruction. In the future we plan to extend our approach to compute also basic block coverage, and then the ``untraceable'' instruction aspect will be fully and consistently eliminated.

Another limitation of our current approach is the limit of 255 registers per method. While this limitation could
potentially affect the success rate of \acvtool, we have encountered only one app, in which this
limit was exceeded after the instrumentation. This limitation can be addressed either by switching to another instrumentation approach, whereby inserting probes as specific method calls, or by splitting big methods. Both of the approaches may require to reassemble an app that has more than 64K methods into a multidex apk~\cite{Multidex}. We plan this extension as the future work.

Our current \acvtool\ prototype does not fully support multidex apps. It is possible to improve the prototype by adding full support for multidex files, as the instrumentation approach itself is extensible to multiple \texttt{dex} files. In our dataset, we have 46 multidex apps, what constitutes 3.5\% of the total population. In particular, in the Google Play benchmark there were 35 apks with 2 \code{dex} files, and 9 apks containing from 3 to 9 \code{dex} files (overall, 44 multidex apps). In the F-Droid benchmark, there were two multidex apps that contained 2 and 42 \code{dex} files, respectively. The current version of \acvtool\ prototype is able to instrument multidex apks and log coverage data for them, but coverage will be reported only for one \code{dex} file. While we considered the multidex apks, if instrumented correctly, as a success for \acvtool, after excluding them, the total instrumentation success rate will become 93.1\%, what is still much higher than other tools.

Also, the current implementation still has a few issues (3.3\% of apps have not survived instrumentation) that we plan to fix in subsequent releases.

\subsection{Threats to validity}
Our experiments with Sapienz reported in Section~\ref{sec:usefulness} allow us to conclude that black-box code coverage measurement provided by \acvtool\ is useful for state-of-art automated testing frameworks. Furthermore, these experiments suggest that different granularities of code coverage could be combined for achieving time-efficient and effective bug finding. 

At this point, it is not yet clear which coverage metric works best. However, the fine-grained instruction-level coverage provided with \acvtool\ has been able to uncover more bugs than other coverage metrics, on our sample. Further investigation of this topic is required to better understand exactly how granularity of code coverage affects the results, and whether there are other confounding factors that may influence the performance of Sapienz and
other similar tools. 

We now discuss the threats to validity for the conclusions we draw from our experiments. These threats to validity could potentially be eliminated by a larger-scale experiment.

\textbf{Internal validity.} Threats to internal validity concern the experiment's aspects that may affect validity of the findings. First, our preliminary experiment involved only a sample of 799 Android apps. It is, in theory, possible that on a larger dataset we will obtain different results in terms of amount of unique crashes and their types. A significantly larger experiment involving thousands of apps could lead to more robust results. 

Second, Sapienz relies on the random input event generator Monkey~\cite{monkey} as the underlying test engine, and thus it is not deterministic. It is possible that this indeterminism may have influence on our current results, and the results obtained with the different coverage metrics could converge on many runs. Our experiments with repeating executions 5 times indicate that this is unlikely. However, the success of Sapienz in finding bugs without coverage information shows that Monkey is powerful enough to explore significant proportions of code, even without evolution of test suites following a coverage-based fitness function. This threat warrants further investigation.

Third, we perform our experiment using the default parameters of Sapienz. It is possible that their values, e.g., the length of a test sequence, may also have an impact on the results. In our future work, we plan to investigate this threat further. 

The last but not least, we acknowledge that the tools measuring code coverage may introduce some additional bugs during the instrumentation process. In our experiments, results for the method and instruction-level coverage have been collected from app versions instrumented with \acvtool, while data for the activity coverage and without coverage were gathered for the original apk versions. If \acvtool\ introduces bugs during instrumentation, this difference may explain why the corresponding populations of crashes for instrumented (method and instruction coverage) and original (activity coverage and no coverage) apps tend to be close. We have tried to address this threat to validity in two ways. First, we have manually inspected a selection of crashes to evaluate whether they appear due to instrumentation. We have not found such evidence. Second, we have run original and instrumented apks with Monkey to assess run-time overhead (as reported in Section~\ref{sec:evaluation}), and we have not seen discrepancies in executions of these apps. If the instrumented version would crash unexpectedly, but the original one would continue running under that same Monkey test, this would be evident from the timings. 

\textbf{External validity.} Threats to external validity concern the generalization of our findings. To test the viability of our hypothesis, we have experimented with only one automated test design tool. It could be possible that other similar tools that rely upon code coverage metrics such as Stoat~\cite{su2017guided}, AimDroid~\cite{gu2017aimdroid} or QBE~\cite{koroglu2018qbe} would not obtain better results when using the fine-grained instruction-level coverage. We plan to investigate this further by extending our experiments to include more automated testing tools that rely on code coverage. 

It should be also stressed that we used apps from the Google Play for our experiment. While preparing a delivery of an app to this market, developers usually apply different post-processing tools, e.g., obfuscators and packers, to prevent potential reverse-engineering. Some crashes in our experiment may be introduced by these tools. In addition, obfuscators may introduce some additional dead code and alter the control flow of apps. These features may also impact the code coverage measurement, especially in case of more fine-grained metrics. Therefore, in our future work we plan also to investigate this issue.

To conclude, our experiment with Sapienz~\cite{mao2016sapienz} has demonstrated that well-known search-based
testing algorithms that rely on code coverage metrics can benefit from the fine-grained code coverage provided by \acvtool. Automated
testing tools may be further improved by including several code coverage metrics with different levels of
measurement granularity. This finding confirms the downstream value of our tool. 

\section{Related work}\label{sec:relwork}
%These frameworks address the needs of the developers (white-box testing and debugging) and the needs of security analysts (black-box testing). Some of these frameworks facilitate manual test creation and replay (e.g., JUnit~\cite{junit}, Mockito~\cite{mockito} and Roboelectric~\cite{roboelectric}), while others target automated test generation. 
\subsection{Android app testing}
%With the increase of Android's popularity, a plethora of frameworks for Android application testing has emerged. We refer an interested reader to the recent surveys in this area. The surveys by Linares et al.~\cite{linares2017continuous} and by Kong et al. \cite{kong2018automated} summarize the challenges and results in mobile testing. Recently, the survey by Choudhary et al.~\cite{choudhary2015automated} has compared the most prominent testing tools that automatically generate app input events in terms of efficiency, including code coverage and fault detection.

%A particular challenge of automated testing on Android is input event generation. Android apps have a complex lifecycle, and they typically have a multitude of entry points. Apps can be triggered by the user, by the Android system, or by fellow third-party apps, and their reaction heavily depends on the context~\cite{arzt2014flowdroid}. Ideally, an automated input generator should be able to find all entry points and all available contexts and generate input sequences to test them. However, in practice it proves challenging.  The available automated testing tools range from random event generators, such as Google's Monkey~\cite{monkey}, to more intricate model-based and code analysis-based frameworks \cite{choudhary2015automated}.  
Automated testing of Android applications is a very prolific research area. Today, there are many frameworks that combine UI and system events generation, striving to achieve better code coverage and fault detection. E.g., Dynodroid~\cite{machiry2013dynodroid} is able to randomly generate both UI and system events. Interaction with the system components via callbacks is another facet, which is addressed by, e.g., EHBDroid~\cite{song2017ehbdroid}. Recently, the survey by Choudhary et al.~\cite{choudhary2015automated} has compared the most prominent testing tools that automatically generate app input events in terms of efficiency, including code coverage and fault detection. Two recent surveys, by Linares et al.~\cite{linares2017continuous} and by Kong et al.~\cite{kong2018automated}, summarize the main efforts and challenges in the automated Android app testing area. 

\subsection{Coverage measurement tools in Android}

\paragraph{White-box coverage measurement}

Tools for white-box code coverage measurement are included into the Android SDK maintained by Google~\cite{google-coverage}.
Supported coverage libraries are JaCoCo~\cite{jacoco}, EMMA~\cite{emma}, and the IntelliJ IDEA coverage
tracker~\cite{intellij}. These tools are capable of measuring fine-grained code coverage, but require that the source
code of an app is available. This makes them suitable only for testing apps at the development stage.

\paragraph{Black-box coverage measurement}

\begin{table*}[ht!]
\centering
\caption{Summary of black-box coverage measuring tools}
\label{tab:relwork}
\setlength\tabcolsep{1.5pt}
\scriptsize
\begin{tabular}{|P{3cm}|P{2cm}|P{2cm}|P{1cm}|P{1.2cm}|P{1.2cm}|P{1.2cm}|P{1.4cm}|P{1cm}|P{1.2cm}|}
%\begin{tabular}{|c|c|c|c|c|c|c|c|c|c|}
\hline
\multirow{3}{3cm}{\centering \textbf{Tool}} &  \multicolumn{3}{c|}{\textbf{Tool details}}& \multicolumn{6}{c|}{\textbf{Results of empirical evaluation}}\\
\cline{2-10}
\cline{2-10}
 & \multirow{2}{2cm}{\centering \textbf{Coverage granularity}} &  \multirow{2}{2cm}{\centering \textbf{Target representation}} & \multirow{2}{1cm}{\centering \textbf{Code available}} &  \multirow{2}{1cm}{\centering \textbf{Sample size}}  & \multicolumn{2}{P{2.4cm}|}{\textbf{Instrumentation success rate (\%)}} & \multicolumn{2}{c|}{\textbf{Overhead}} & \multirow{2}{1.2cm}{\centering \textbf{Compli- ance evaluated}} \\
%& &  &  & &  &  &  &  &  \\
\cline{6-9}
 &  &   & &  & \textbf{Instru- mented}& \textbf{Executed}&  \textbf{Instr. time (sec/app)} & \textbf{Run time (\%)} &   \\
%   & & &  &  &   &  &  \textbf{(sec/app)} &  \textbf{(\%)}  & \\
%      & &  &  &  &  &    &   & \\

\hline
\hline 
%\multicolumn{10}{|c|}{\textbf{Benign Applications}} \\ 
%\hline
ELLA~\cite{ella,wang2018empirical} & method & Dalvik bytecode & Y & 68 \cite{wang2018empirical} & 60\% \cite{wang2018empirical} & 60\% \cite{wang2018empirical}  & N/A & N/A & N/A \\ 
%  & &  &  &  &  &  &  &  &\\
\hline
Huang et al.~\cite{huang2015code} & class, method, basic block, instruction & Dalvik bytecode (\code{smali}) & N & 90 & 36\% & N/A & N/A   & N/A  & Y\\ 
% & method, & & &  &  &  &  &   & \\
\hline
BBoxTester~\cite{zhauniarovich2015towards} & class, method, basic block  & Java bytecode & Y & 91 & 65\% & N/A & 15.5  & N/A &  N\\ 
%   & &  & &  &  &  & &  & \\
\hline
Asc~\cite{song2017ehbdroid} & basic block, instruction & Jimple &Y & 35 & N/A & N/A &  N/A& N/A & N \\
%   &  & & &  &  &  & &  & \\ 

\hline
InsDal~\cite{liu2017insdal,yan2016target,lu2016lightweight} & class, method & Dalvik bytecode & N & 10 & N/A & N/A & 1.5  & N/A & N \\ 
\hline
Sapienz~\cite{mao2016sapienz} & activity & Dalvik bytecode & Y & 1112 & N/A  & N/A & N/A & N/A & N\\ 
\hline 
DroidFax~\cite{cai2017droidfax,cai2017droidcat,cai2017understanding} & instruction & Jimple & Y & 195 &  N/A & N/A & N/A & N/A & N \\ 
\hline
AndroCov~\cite{borges2018guiding,androcov} & method, instruction & Jimple & Y & 17 &  N/A & N/A & N/A & N/A & N \\ 
\hline
CovDroid~\cite{yeh2015covdroid} & method & Dalvik bytecode (\code{smali}) & N & 1 &  N/A & N/A & N/A & N/A & N \\
\hline
\hline 
\textbf{\acvtool} (this paper) & \textbf{class, method, instruction}  & \textbf{Dalvik bytecode (\code{smali})} & \textbf{Y} & \textbf{1278}  & \textbf{97.8\%} & \textbf{96.9\%}  & \textbf{33.3}  & up to \textbf{27\%} on PassMark & \textbf{Y} \\ 
\hline
\end{tabular}
\end{table*}

Several frameworks for measuring black-box code coverage of Android apps already exist, however they are inferior to
\acvtool. Notably, these frameworks often measure code coverage at coarser granularity. For example, ELLA~\cite{ella}, InsDal~\cite{liu2017insdal}, and CovDroid~\cite{yeh2015covdroid} measure code coverage only at the method level.

ELLA~\cite{ella} is arguably one of the most popular tools to measure Android code coverage in the black-box setting, however, it is no longer supported. An empirical study by Wang et al.~\cite{wang2018empirical} has evaluated performance of Monkey~\cite{monkey}, Sapienz~\cite{mao2016sapienz}, Stoat~\cite{su2017guided}, and WCTester~\cite{zeng2016automated} automated testing tools on large and popular industry-scale apps, such as Facebook, Instagram and Google. They have used ELLA to measure method code coverage, and they reported the total success rate of ELLA at 60\% (41 apps) on their sample of 68 apps.

%Their results show that it a combination of Monkey and Sapienz is the most effective in achieving high code coverage and finding faults. Wang et al.~\cite{wang2018empirical} also 

Huang et al.~\cite{huang2015code} proposed an approach to measure code coverage for dynamic analysis tools for
Android apps. Their high-level approach is similar to ours: an app is decompiled into \texttt{smali} files, and these files are
instrumented by placing probes at every class, method and basic block to track their execution. However, the authors
report a low instrumentation success rate of 36\%, and only 90 apps have been used for evaluation.  Unfortunately, the tool is not
publicly available, and we were unable to obtain it or the dataset by contacting the authors. Because of this, we cannot compare its
performance with \acvtool, although we report much higher instrumentation rate, evaluated against much larger dataset.

BBoxTester~\cite{zhauniarovich2015towards} is another tool for measuring black-box code coverage. Its workflow includes app disassembling with \texttt{apktool} and decompilation of the \texttt{dex} files into Java \code{jar} files using \texttt{dex2jar}~\cite{dex2jar}. The \code{jar} files are instrumented using EMMA~\cite{emma}, and assembled back into an apk. The empirical evaluation of BBoxTester showed the successful repackaging rate of 65\%, and the instrumentation time has been reported to be 15 seconds per app. We were able to obtain the original BBoxTester dataset. Out of 91 apps, \acvtool\ failed to instrument just one. This error was not due to our own instrumentation code: \texttt{apktool} could not repackage this app. Therefore, \acvtool\ successfully instrumented 99\% of this dataset, against 65\% of BBoxTester.

The InsDal tool~\cite{liu2017insdal} instruments apps for class and method-level coverage logging by inserting probes in the \texttt{smali} code, and its workflow is similar to \acvtool.
The tool has been applied for measuring code coverage in the black-box setting with the AppTag tool~\cite{yan2016target}, and for logging the
number of method invocations in measuring the energy consumption of apps~\cite{lu2016lightweight}. The information about instrumentation success rate is not available for InsDal, and it has been evaluated on a limited dataset of 10 apps. The authors have reported average instrumentation time overhead of 1.5 sec per app, and average instrumentation code overhead of 18.2\% of \texttt{dex} file size. \acvtool\ introduces smaller code size overhead of 11\%, on average, but requires more time to instrument an app. On our dataset, average instrumentation time is 24.1 seconds per app, when instrumenting at the method level only. It is worth noting that half of this time is spent on repackaging with \texttt{apktool}.

%(when instrumenting at the method level, similarly to InsDal), but requires more time to instrument an app (on our dataset, 24.1 seconds at the method level and 33.3 seconds at the instruction level per app on average
%Yan et al.~\cite{yan2016target} report the experimental results for five open-source apps that have been instrumented with both InsDal~\cite{liu2017insdal} and EMMA~\cite{emma}. However, the authors~\cite{yan2016target} did not specify if they used the same test sequence for both tools. Thus, it is difficult to conclude whether the compliance of the code coverage measurement has been evaluated.  

CovDroid~\cite{yeh2015covdroid}, another black-box code coverage measurement system for Android apps, transforms apk code into \texttt{smali}-representation using the \texttt{smali} disassembler~\cite{smali} and inserts probes at the method level. The coverage data is collected using an execution monitor, and the tool is able to collect timestamps for executed methods. While the instrumentation approach of \acvtool\ is similar in nature to that of CovDroid, the latter tool has been evaluated on a single application only. 

%EHBDroid~\cite{song2017ehbdroid} relies upon a library called \texttt{Asc} to measure code coverage in basic blocks and
%statements in the \texttt{Jimple} code representation. The authors~\cite{song2017ehbdroid} have not provided
%information related to the success rate and the performance of \texttt{Asc}. 

%The DroidFax~\cite{cai2017droidfax} toolbox for Android app analysis integrates a black-box code coverage tracker at the
%basic block, method and class level. Cai and Ryder~\cite{cai2017understanding}, and Cai et al.~\cite{cai2017droidcat}
%report on using this toolbox, but they do not mention instrumentation success rate statistics.

%\paragraph{Instrumentation}

Alternative approaches to Dalvik instrumentation focus on performing detours via other languages, e.g., Java or Jimple.
For example, Bartel et al.~\cite{bartel2012improving} worked on instrumenting Android apps for improving their privacy
and security via translation to Java bytecode. Zhauniarovich et al.~\cite{zhauniarovich2015towards} translated Dalvik
into Java bytecode in order to use EMMA's code coverage measurement functionality. However, the limitation of such
approaches, as reported in~\cite{zhauniarovich2015towards}, is that not all apps can be retargeted into Java bytecode.

The instrumentation of apps translated into the Jimple representation has been used in, e.g.,
Asc~\cite{song2017ehbdroid}, DroidFax~\cite{cai2017droidfax}, and AndroCov~\cite{androcov,borges2018guiding}. Jimple is a suitable representation for subsequent analysis with Soot~\cite{arzt2017soot},  yet, unlike \texttt{smali}, it does not belong to the ``core'' Android
technologies maintained by Google. Moreover, Arnatovich et al.~\cite{arnatovich2014empirical} in their comparison of different intermediate representations for Dalvik bytecode advocate that \texttt{smali} is the most accurate alternative to the original Java source code and therefore is the most suitable for security testing.

Remarkably, in the absence of reliable fine-grained code coverage reporting tools, some frameworks~\cite{mao2016sapienz,song2017ehbdroid,cai2017droidfax,li2017droidbot,carter2016curiousdroid} integrate their own black-box coverage measurement libraries. Many of these papers do note that they have to design their own code coverage measurement means in the absence of a reliable tool. \acvtool\ addresses this need of the community. As the coverage measurement is not the core contribution of these works, the authors have not provided enough information about the rates of successful instrumentation, and other details related to the performance of these libraries, so we are not able to compare them with \acvtool. %Yet, the very existence of these plugins demonstrates that there is a huge need in the community for reliable coverage measurement tools like \acvtool.

%Logging method stack trace information has been used to evaluate method coverage in DroidBot~\cite{li2017droidbot} and CuriousDroid~\cite{carter2016curiousdroid}. 
\paragraph{App instrumentation}
Among the Android application instrumentation approaches, the most relevant for us are the techniques discussed by
Huang et al.~\cite{huang2015code}, InsDal~\cite{liu2017insdal} and CovDroid~\cite{yeh2015covdroid}. \acvtool\ shows much better instrumentation success rate, because our instrumentation approach deals with many peculiarities of the Dalvik bytecode. A similar instrumentation approach has been also used in the DroidLogger~\cite{dai2012droidlogger} and SwiftHand~\cite{choi2013guided} frameworks, which do not report their instrumentation success rates.

\paragraph{Summary}
Table~\ref{tab:relwork} summarizes the performance of \acvtool\ and code coverage granularities that it supports in
comparison to other state-of-the-art tools. \acvtool\ significantly outperforms any other tool that measures black-box
code coverage of Android apps. Our tool has been extensively tested with real-life applications, and it has excellent
instrumentation success rate, in contrast to other tools, e.g.,~\cite{huang2015code} and~\cite{zhauniarovich2015towards}. We attribute the reliable performance of \acvtool\ to the very detailed investigation
of \texttt{smali} instructions we have done, that is missing in the literature. \acvtool\ is available as
open-source to share our insights with the community, and to replace the outdated tools (ELLA~\cite{ella} and
BBoxTester\cite{zhauniarovich2015towards}) or publicly unavailable tools (\cite{huang2015code,yeh2015covdroid}).

%\Ans{We see the novelty in the following aspects: 1) ACVTool is the first tool for fine-grained code coverage measurement that works with the majority of third-party apks. ACVTool integrates several novel smali instrumentation solutions, e.g., correct probe insertion in case of move-result* and other instructions. Other papers, e.g., [21], do not report these details, their code is not available, and report lower success rate. Empirical comparison with [21] is impossible because the tool was not released, and the paper lacks details for replicating it. 2) We have performed a large-scale evaluation of ACVTool using far more apps than other tools, including Google Play dataset. 3) We give evidence that fine-grained code coverage can be useful per se in automated testing tools. This opens new possibilities for further improvements of automated Android testing.}

\section{Conclusions}\label{sec:conclusions}

In this paper, we presented an instrumentation technique for Android apps. We incorporated this technique into \acvtool\
-- an effective and efficient tool for measuring precise code coverage of Android apps. We were able to instrument and
execute 96.9\% out of 1278 apps used for the evaluation, showing that \acvtool\ is practical and reliable.

The empirical evaluation that we have performed allows us to conclude that \acvtool\ will be useful for both researchers
who are building testing, program analysis, and security assessment tools for Android, and practitioners in industry who
need reliable and accurate coverage information.

To enable better support for automated testing community, we are working to add support for multidex apps, extend the set of available coverage metrics
to branch coverage, and to alleviate the limitation caused by the fixed amount of registers in a method. Also, as an interesting line of future work, we consider on-the-fly \texttt{dex} file instrumentation that will make
\acvtool\ even more useful in the context of analyzing highly complex applications and malware. 

Furthermore, our experiments with Sapienz have produced interesting conclusions that the granularity of coverage is important, when used as a component of the fitness function in the black-box app testing. The second line of the future work for us is to expand our experiments to more apps and more testing tools, thus establishing better guidelines on which coverage metric is more effective and efficient in bug finding.

\subsection*{Acknowledgements}
This work has been partially supported by Luxembourg National Research Fund through grants C15/IS/10404933/COMMA and AFR-PhD-11289380-DroidMod.

\bibliographystyle{plain}
%\bibliography{references_full}

\begin{thebibliography}{10}

\bibitem{dex2jar}
dex2jar, 2017.

\bibitem{allix2016androzoo}
K.~Allix, T.~F. Bissyande, J.~Klein, and Y.~L. Traon.
\newblock Androzoo: Collecting millions of android apps for the research
  community.
\newblock In {\em 2016 IEEE/ACM 13th Working Conference on Mining Software
  Repositories (MSR)}, pages 468--471, May 2016.

\bibitem{ammann2016introduction}
Paul Ammann and Jeff Offutt.
\newblock {\em Introduction to Software Testing}.
\newblock Cambridge University Press, 2 edition, 2016.

\bibitem{arnatovich2014empirical}
Yauhen~Leanidavich Arnatovich, Hee Beng~Kuan Tan, and Lwin~Khin Shar.
\newblock Empirical comparison of intermediate representations for android
  applications.
\newblock In {\em 26th International Conference on Software Engineering and
  Knowledge Engineering}, 2014.

\bibitem{arzt2017soot}
Steven Arzt, Siegfried Rasthofer, and Eric Bodden.
\newblock The soot-based toolchain for analyzing android apps.
\newblock In {\em Proceedings of the 4th International Conference on Mobile
  Software Engineering and Systems}, MOBILESoft '17, pages 13--24, Piscataway,
  NJ, USA, 2017. IEEE Press.

\bibitem{azim2013targeted}
Tanzirul Azim and Iulian Neamtiu.
\newblock Targeted and depth-first exploration for systematic testing of
  android apps.
\newblock In {\em Proceedings of the 2013 ACM SIGPLAN International Conference
  on Object Oriented Programming Systems Languages \&\#38; Applications},
  OOPSLA '13, pages 641--660, New York, NY, USA, 2013. ACM.

\bibitem{backes2017artist}
M.~Backes, S.~Bugiel, O.~Schranz, P.~v.~Styp-Rekowsky, and S.~Weisgerber.
\newblock Artist: The android runtime instrumentation and security toolkit.
\newblock In {\em 2017 IEEE European Symposium on Security and Privacy (EuroS
  P)}, pages 481--495, April 2017.

\bibitem{bartel2012improving}
Alexandre Bartel, Jacques Klein, Martin Monperrus, Kevin Allix, and Yves~Le
  Traon.
\newblock In-vivo bytecode instrumentation for improving privacy on android
  smartphones in uncertain environments, 2012.

\bibitem{borges2018guiding}
Nataniel~P. Borges, Jr., Maria G\'{o}mez, and Andreas Zeller.
\newblock Guiding app testing with mined interaction models.
\newblock In {\em Proceedings of the 5th International Conference on Mobile
  Software Engineering and Systems}, MOBILESoft '18, pages 133--143, New York,
  NY, USA, 2018. ACM.

\bibitem{DvmInternals_Bornstein2008}
D.~Bornstein.
\newblock {Google I/O 2008 - Dalvik Virtual Machine Internals}, 2008.

\bibitem{cai2017droidcat}
H.~Cai, N.~Meng, B.~Ryder, and D.~Yao.
\newblock Droidcat: Effective android malware detection and categorization via
  app-level profiling.
\newblock {\em IEEE Transactions on Information Forensics and Security}, pages
  1--1, 2018.

\bibitem{cai2017droidfax}
H.~Cai and B.~G. Ryder.
\newblock Droidfax: A toolkit for systematic characterization of android
  applications.
\newblock In {\em 2017 IEEE International Conference on Software Maintenance
  and Evolution (ICSME)}, pages 643--647, Sep. 2017.

\bibitem{cai2017understanding}
H.~Cai and B.~G. Ryder.
\newblock Understanding android application programming and security: A dynamic
  study.
\newblock In {\em 2017 IEEE International Conference on Software Maintenance
  and Evolution (ICSME)}, pages 364--375, Sep. 2017.

\bibitem{carter2016curiousdroid}
Patrick Carter, Collin Mulliner, Martina Lindorfer, William Robertson, and
  Engin Kirda.
\newblock Curiousdroid: Automated user interface interaction for android
  application analysis sandboxes.
\newblock In Jens Grossklags and Bart Preneel, editors, {\em Financial
  Cryptography and Data Security}, pages 231--249, Berlin, Heidelberg, 2017.
  Springer Berlin Heidelberg.

\bibitem{chekam2017empirical}
Thierry~Titcheu Chekam, Mike Papadakis, Yves~Le Traon, and Mark Harman.
\newblock An empirical study on mutation, statement and branch coverage fault
  revelation that avoids the unreliable clean program assumption.
\newblock In {\em Proceedings of the 39th International Conference on Software
  Engineering}, ICSE '17, pages 597--608, Piscataway, NJ, USA, 2017. IEEE
  Press.

\bibitem{choi2013guided}
Wontae Choi, George Necula, and Koushik Sen.
\newblock Guided gui testing of android apps with minimal restart and
  approximate learning.
\newblock In {\em Proceedings of the 2013 ACM SIGPLAN International Conference
  on Object Oriented Programming Systems Languages \&\#38; Applications},
  OOPSLA '13, pages 623--640, New York, NY, USA, 2013. ACM.

\bibitem{choudhary2015automated}
Shauvik~Roy Choudhary, Alessandra Gorla, and Alessandro Orso.
\newblock Automated test input generation for android: Are we there yet? (e).
\newblock In {\em Proceedings of the 2015 30th IEEE/ACM International
  Conference on Automated Software Engineering (ASE)}, ASE '15, pages 429--440,
  Washington, DC, USA, 2015. IEEE Computer Society.

\bibitem{KotlinAnnouncement}
Mike Cleron.
\newblock {Android Announces Support for Kotlin}, May 2017.

\bibitem{dai2012droidlogger}
Shuaifu Dai, Tao Wei, and Wei Zou.
\newblock Droidlogger: Reveal suspicious behavior of android applications via
  instrumentation.
\newblock In {\em 2012 7th International Conference on Computing and
  Convergence Technology (ICCCT)}, pages 550--555, Dec 2012.

\bibitem{dashevskyi2018influence}
Stanislav Dashevskyi, Olga Gadyatskaya, Aleksandr Pilgun, and Yury
  Zhauniarovich.
\newblock The influence of code coverage metrics on automated testing
  efficiency in android.
\newblock In {\em Proceedings of the 2018 ACM SIGSAC Conference on Computer and
  Communications Security}, CCS '18, pages 2216--2218, New York, NY, USA, 2018.
  ACM.

\bibitem{ella}
{ELLA}.
\newblock A tool for binary instrumentation of {Android} apps, 2016.

\bibitem{gligoric2015guidelines}
Milos Gligoric, Alex Groce, Chaoqiang Zhang, Rohan Sharma, Mohammad~Amin
  Alipour, and Darko Marinov.
\newblock Guidelines for coverage-based comparisons of non-adequate test
  suites.
\newblock {\em ACM Trans. Softw. Eng. Methodol.}, 24(4):22:1--22:33, September
  2015.

\bibitem{Multidex}
{Google}.
\newblock {Enable Multidex for Apps with Over 64K Methods}.

\bibitem{monkey}
Google.
\newblock {UI/Application Exerciser Monkey}.

\bibitem{DexFormat}
{Google}.
\newblock {Dalvik Executable format}, 2017.

\bibitem{dalvik}
{Google}.
\newblock Dalvik bytecode, 2018.

\bibitem{google-smali}
Google.
\newblock smali, 2018.

\bibitem{google-coverage}
Google.
\newblock Test your app, 2018.

\bibitem{gu2017aimdroid}
T.~Gu, C.~Cao, T.~Liu, C.~Sun, J.~Deng, X.~Ma, and J.~Lü.
\newblock Aimdroid: Activity-insulated multi-level automated testing for
  android applications.
\newblock In {\em 2017 IEEE International Conference on Software Maintenance
  and Evolution (ICSME)}, pages 103--114, Sep. 2017.

\bibitem{huang2015code}
C.~Huang, C.~Chiu, C.~Lin, and H.~Tzeng.
\newblock Code coverage measurement for android dynamic analysis tools.
\newblock In {\em 2015 IEEE International Conference on Mobile Services}, pages
  209--216, June 2015.

\bibitem{jacoco}
JaCoCo.
\newblock Java code coverage library, 2018.

\bibitem{smali}
{JesusFreke}.
\newblock smali/backsmali.

\bibitem{intellij}
JetBrains.
\newblock Code coverage, 2017.

\bibitem{kong2018automated}
P.~Kong, L.~Li, J.~Gao, K.~Liu, T.~F. Bissyande, and J.~Klein.
\newblock Automated testing of android apps: A systematic literature review.
\newblock {\em IEEE Transactions on Reliability}, pages 1--22, 2018.

\bibitem{koroglu2018qbe}
Y.~Koroglu, A.~Sen, O.~Muslu, Y.~Mete, C.~Ulker, T.~Tanriverdi, and Y.~Donmez.
\newblock Qbe: Qlearning-based exploration of android applications.
\newblock In {\em 2018 IEEE 11th International Conference on Software Testing,
  Verification and Validation (ICST)}, pages 105--115, April 2018.

\bibitem{li2013bytecode}
N.~Li, X.~Meng, J.~Offutt, and L.~Deng.
\newblock Is bytecode instrumentation as good as source code instrumentation:
  An empirical study with industrial tools (experience report).
\newblock In {\em 2013 IEEE 24th International Symposium on Software
  Reliability Engineering (ISSRE)}, pages 380--389, Nov 2013.

\bibitem{androcov}
Y.~Li.
\newblock {AndroCov}. measure test coverage without source code, 2016.

\bibitem{li2017droidbot}
Yuanchun Li, Ziyue Yang, Yao Guo, and Xiangqun Chen.
\newblock Droidbot: a lightweight ui-guided test input generator for android.
\newblock In {\em 2017 IEEE/ACM 39th International Conference on Software
  Engineering Companion (ICSE-C)}, pages 23--26, May 2017.

\bibitem{linares2017continuous}
M.~Linares-Vásquez, K.~Moran, and D.~Poshyvanyk.
\newblock Continuous, evolutionary and large-scale: A new perspective for
  automated mobile app testing.
\newblock In {\em 2017 IEEE International Conference on Software Maintenance
  and Evolution (ICSME)}, pages 399--410, Sep. 2017.

\bibitem{liu2017insdal}
J.~Liu, T.~Wu, X.~Deng, J.~Yan, and J.~Zhang.
\newblock Insdal: A safe and extensible instrumentation tool on dalvik
  byte-code for android applications.
\newblock In {\em 2017 IEEE 24th International Conference on Software Analysis,
  Evolution and Reengineering (SANER)}, pages 502--506, Feb 2017.

\bibitem{lu2016lightweight}
Q.~Lu, T.~Wu, J.~Yan, J.~Yan, F.~Ma, and F.~Zhang.
\newblock Lightweight method-level energy consumption estimation for android
  applications.
\newblock In {\em 2016 10th International Symposium on Theoretical Aspects of
  Software Engineering (TASE)}, pages 144--151, July 2016.

\bibitem{machiry2013dynodroid}
Aravind Machiry, Rohan Tahiliani, and Mayur Naik.
\newblock Dynodroid: An input generation system for android apps.
\newblock In {\em Proceedings of the 2013 9th Joint Meeting on Foundations of
  Software Engineering}, ESEC/FSE 2013, pages 224--234, New York, NY, USA,
  2013. ACM.

\bibitem{mao2016sapienz}
Ke~Mao, Mark Harman, and Yue Jia.
\newblock Sapienz: Multi-objective automated testing for android applications.
\newblock In {\em Proceedings of the 25th International Symposium on Software
  Testing and Analysis}, ISSTA 2016, pages 94--105, New York, NY, USA, 2016.
  ACM.

\bibitem{pilgun2018effective}
Aleksandr Pilgun, Olga Gadyatskaya, Stanislav Dashevskyi, Yury Zhauniarovich,
  and Artsiom Kushniarou.
\newblock An effective android code coverage tool.
\newblock In {\em Proceedings of the 2018 ACM SIGSAC Conference on Computer and
  Communications Security}, CCS '18, pages 2189--2191, New York, NY, USA, 2018.
  ACM.

\bibitem{emma}
V.~Rubtsov.
\newblock Emma: Java code coverage tool, 2006.

\bibitem{sadeghi2017patdroid}
Alireza Sadeghi, Reyhaneh Jabbarvand, and Sam Malek.
\newblock Patdroid: Permission-aware gui testing of android.
\newblock In {\em Proceedings of the 2017 11th Joint Meeting on Foundations of
  Software Engineering}, ESEC/FSE 2017, pages 220--232, New York, NY, USA,
  2017. ACM.

\bibitem{passmark}
PassMark Software.
\newblock {Passmark}. interpreting your results from performancetest, 2018.

\bibitem{song2017ehbdroid}
Wei Song, Xiangxing Qian, and Jeff Huang.
\newblock Ehbdroid: Beyond gui testing for android applications.
\newblock In {\em Proceedings of the 32Nd IEEE/ACM International Conference on
  Automated Software Engineering}, ASE 2017, pages 27--37, Piscataway, NJ, USA,
  2017. IEEE Press.

\bibitem{su2017guided}
Ting Su, Guozhu Meng, Yuting Chen, Ke~Wu, Weiming Yang, Yao Yao, Geguang Pu,
  Yang Liu, and Zhendong Su.
\newblock Guided, stochastic model-based gui testing of android apps.
\newblock In {\em Proceedings of the 2017 11th Joint Meeting on Foundations of
  Software Engineering}, ESEC/FSE 2017, pages 245--256, New York, NY, USA,
  2017. ACM.

\bibitem{tam2015copperdroid}
K.~Tam, S.~Khan, A.~Fattori, and L.~Cavallaro.
\newblock {CopperDroid: Automatic} reconstruction of {Android} malware
  behaviors.
\newblock In {\em Proceedings of the Network and Distributed System Security
  Symposium (NDSS)}, 2015.

\bibitem{tengeri2016negative}
D.~Tengeri, F.~Horváth, Á. Beszédes, T.~Gergely, and T.~Gyimóthy.
\newblock Negative effects of bytecode instrumentation on java source code
  coverage.
\newblock In {\em 2016 IEEE 23rd International Conference on Software Analysis,
  Evolution, and Reengineering (SANER)}, volume~1, pages 225--235, March 2016.

\bibitem{jimple}
R.~Vallee-rai and L.~Hendren.
\newblock Jimple: {Simplifying Java} bytecode for analyses and transformations.
\newblock 2004.

\bibitem{wang2018empirical}
Wenyu Wang, Dengfeng Li, Wei Yang, Yurui Cao, Zhenwen Zhang, Yuetang Deng, and
  Tao Xie.
\newblock An empirical study of android test generation tools in industrial
  cases.
\newblock In {\em Proceedings of the 33rd ACM/IEEE International Conference on
  Automated Software Engineering}, ASE 2018, pages 738--748, New York, NY, USA,
  2018. ACM.

\bibitem{Apktool_webpage}
R.~Wi\'{s}niewski and C.~Tumbleson.
\newblock {Apktool - A} tool for reverse engineering 3rd party, closed, binary
  {Android} apps, 2017.

\bibitem{wohlin2012experimentation}
Claes Wohlin, Per Runeson, Martin Hst, Magnus~C. Ohlsson, Bjrn Regnell, and
  Anders Wessln.
\newblock {\em Experimentation in Software Engineering}.
\newblock Springer Publishing Company, Incorporated, 2012.

\bibitem{wong2016intellidroid}
M.~Y Wong and D.~Lie.
\newblock {IntelliDroid: A} targeted input generator for the dynamic analysis
  of {Android} malware.
\newblock In {\em Proceedings of the Network and Distributed System Security
  Symposium (NDSS)}, 2016.

\bibitem{yan2016target}
Jiwei Yan, Tianyong Wu, Jun Yan, and Jian Zhang.
\newblock Target directed event sequence generation for android applications,
  2016.

\bibitem{Apkil_webpage}
K.~Yang.
\newblock {APK Instrumentation} library, 2018.

\bibitem{yang2009survey}
Q.~Yang, J.~J. Li, and D.~M. Weiss.
\newblock A survey of coverage-based testing tools.
\newblock {\em The Computer Journal}, 52(5):589--597, Aug 2009.

\bibitem{yeh2015covdroid}
C.~Yeh and S.~Huang.
\newblock Covdroid: A black-box testing coverage system for android.
\newblock In {\em 2015 IEEE 39th Annual Computer Software and Applications
  Conference}, volume~3, pages 447--452, July 2015.

\bibitem{yoo2012regression}
S.~Yoo and M.~Harman.
\newblock Regression testing minimization, selection and prioritization: A
  survey.
\newblock {\em Softw. Test. Verif. Reliab.}, 22(2):67--120, March 2012.

\bibitem{zeng2016automated}
Xia Zeng, Dengfeng Li, Wujie Zheng, Fan Xia, Yuetang Deng, Wing Lam, Wei Yang,
  and Tao Xie.
\newblock Automated test input generation for android: Are we really there yet
  in an industrial case?
\newblock In {\em Proceedings of the 2016 24th ACM SIGSOFT International
  Symposium on Foundations of Software Engineering}, FSE 2016, pages 987--992,
  New York, NY, USA, 2016. ACM.

\bibitem{zhauniarovich2014fsquadra}
Yury Zhauniarovich, Olga Gadyatskaya, Bruno Crispo, Francesco La~Spina, and
  Ermanno Moser.
\newblock Fsquadra: fast detection of repackaged applications.
\newblock In {\em IFIP Annual Conference on Data and Applications Security and
  Privacy}, pages 130--145. Springer, 2014.

\bibitem{zhauniarovich2015towards}
Yury Zhauniarovich, Anton Philippov, Olga Gadyatskaya, Bruno Crispo, and Fabio
  Massacci.
\newblock {Towards Black Box Testing of Android Apps}.
\newblock In {\em 2015 Tenth International Conference on Availability,
  Reliability and Security}, pages 501--510, August 2015.

\end{thebibliography}

\end{document}